\documentstyle[11pt,aaspp4,tighten]{article}
\def\deg{\ifmmode^\circ\else$^\circ$\fi}

\begin{document}
\title{\bf POLARIMETRY AND UNIFICATION OF LOW-REDSHIFT RADIO GALAXIES}

\author{Marshall H. Cohen\altaffilmark{1}, Patrick M. Ogle\altaffilmark{1,2}, 
Hien D. Tran\altaffilmark{3,4}}

\altaffiltext{1}{California Institute of Technology, Pasadena, CA 91125; 
    mhc@astro.caltech.edu}
\altaffiltext{2}{present address: Massachusetts Institute of Technology;
    pmo@space.mit.edu}
\altaffiltext{3}{Institute of Geophysics and Planetary Physics, Lawrence Livermore 
    National Laboratory, 7000 East Ave, P.O. Box 808, L413, Livermore, CA 94550}
\altaffiltext{4}{present address: Johns Hopkins University, Baltimore, MD 21218; 
    tran@adcam.pha.jhu.edu}

\author{Robert W. Goodrich\altaffilmark{5}, Joseph S. Miller\altaffilmark{6}}

\altaffiltext{5}{CARA/Keck Observatory, 65-1120 Mamalahoa Highway, Kamuela, HI 96743; 
    goodrich@keck.hawaii.edu}
\altaffiltext{6}{UCO/Lick Observatory, University of California, Santa Cruz, CA 95064;
    miller@ucolick.org}

\begin{abstract}

   We have made high-quality measurements of the polarization spectra
of 13 FR II radio galaxies and taken polarization images for 11 of
these with the Keck telescopes. Seven of the eight narrow-line radio
galaxies (NLRG) are polarized, and six of the seven show prominent
broad Balmer lines in polarized light. The broad lines are also weakly
visible in total flux. Some of the NLRG show bipolar regions with
roughly circumferential polarization vectors, revealing a large
reflection nebula illuminated by a central source.  Our observations
powerfully support the hidden quasar hypothesis for some NLRG.
According to this hypothesis, the continuum and broad lines are blocked
by a dusty molecular torus, but can be seen by reflected, hence
polarized, light. Classification as NLRG, broad-line radio galaxy
(BLRG), or quasar therefore depends on orientation. However, not all
objects fit into this unification scheme. Our sample is biased towards
objects known in advance to be polarized, but the combination of our
results with those of Hill, Goodrich, \& DePoy (1996) show that at
least 6 out of a complete, volume and flux-limited sample of 9 FR II
NLRG have broad lines, seen either in polarization or P$\alpha$.

   The BLRG in our sample range from 3C 382, which has a quasar-like
spectrum, to the highly-reddened IRAS source FSC 2217+259.  This
reddening sequence suggests a continuous transition from unobscured
quasar to reddened BLRG to NLRG. Apparently the obscuring torus does
not have a distinct edge.  The BLRG have polarization images which are
consistent with a point source broadened by seeing and diluted by
starlight. We do not detect extended nebular or scattered emission,
perhaps because it is swamped by the nuclear source.  Our
starlight-corrected BLRG spectra can be explained with a two-component
model: a quasar viewed through dust, and quasar light scattered by
dust.  The direct flux is more reddened than the scattered flux,
causing the polarization to rise steeply to the blue. Strong rotations
of the electric vector position angle across H$\alpha$ in 3C 227 and 3C
445 may be explained by systematic orbital motions in an equatorial
broad-line region.

\end{abstract}


\section{Introduction}          

  Active galaxies can be classified in many ways, according to
spectrum, luminosity, and morphology, and the result is a confused
taxonomy that only partly reflects the underlying physics. This
situation has led to ``unification theories" in which various
categories are linked by orientation, evolution or a gradient in
physical characteristics.  (For reviews, see e.g.
\cite{law87,urr95}).  The one that concerns us in this paper connects
Type 1 (broad-line) and Type 2 (narrow-line) objects by aspect. In this
picture an active galactic nucleus (AGN) contains a bright continuum
source and a broad-line region (BLR) surrounded by a dusty torus. When
viewed from near the pole the center is visible and a Type 1 object is
seen, whereas along the equator the view of the center is blocked by
the torus, and a Type 2 object is seen.  When the conditions are right
the nuclear light (continuum and broad lines) escaping along the poles
can be scattered and thus polarized, and seen by the observer.

   Miller and his collaborators (\cite{ant85,mil90,tra92}), with the
aid of spectropolarimetry, showed that many Seyfert 2 galaxies do
contain a hidden BLR, and a detailed investigation of the Seyfert 2 NGC
1068 (\cite{mil91}) showed that the scattered light is typical of a
Seyfert 1 nucleus. Thus unification by aspect is proven for many
Seyfert 1 and 2 galaxies. \cite{hei97} surveyed Seyfert 2's selected by
FIR fluxes and colors, which supposedly are isotropic quantities, and
showed that at least 7 of 16 objects contain a hidden BLR. Also,
infrared observations that penetrate the dust and see the BLR have
shown directly that some Type 2 Seyferts have broad line regions
(\cite{vei97}).

  In this paper our interest centers on the powerful FR II radio
galaxies (RG), and on radio-loud QSO's (quasars).  Like Seyferts, the
RG come in Types 1 and 2 (BLRG and NLRG, respectively), and there has
been interest in seeing if they also can be unified by aspect. They
have the further virtue of containing a radio jet that presumably marks
the dynamical axis of the black hole and accretion disk, providing
additional geometric information.  Quasars and RG are obviously
anisotropic at radio wavelengths, and the viewing aspect has long been
a major factor in their unification (e.g.. \cite{rea78}).  The common
deficit of observed ionizing photons relative to those seen by the
line-emitting regions (e.g. \cite{neu80,bin93}) shows that RG also must
be anisotropic at optical and UV wavelengths.  A statistical basis for
unifying RG and quasars was provided by \cite{bar89}, who used the
opaque torus scenario developed for Seyferts. Barthel postulated that
most FR II RG and quasars are similar, and contain a continuum source
and  BLR inside an opaque torus.  Quasars are seen within about 45
degrees of the pole where there is a direct view into the center, and
RG are seen at higher inclinations where the quasar is partially or
entirely hidden.  Blazars are seen when the line of sight (LOS) is
close to the axis, and the rapid variability in flux and polarization
due to variations in beamed synchrotron radiation can be seen.

   Many observations have now shown that much of this scenario must be
correct. Broad H$\alpha$ has been seen in the polarized light from the
NLRG 3C 234 (e.g. \cite{tra95b}, hereafter T95), 3C 321 (\cite{you96}),
and Cygnus A (\cite{ogl97}, hereafter O97).  In all three objects an
estimate of luminosity is consistent with their being off-axis
quasars.  Imaging polarimetry shows that in some RG the polarized
regions have the azimuthal symmetry of a reflection nebula, as expected
from an extended scattering region illuminated by a point source
(\cite{dis93,dra93a,coh96}, O97, \cite{tra98}).  Also, as with
Seyferts, broad IR lines have been seen in a number of NLRG.
\cite{hil96} found broad P$\alpha$ in 4 of 9 NLRG that
they surveyed, indicating that some NLRG contain a BLR hidden by dust,
and would be called quasars if seen from a dust-free direction.

   Another aspect of unification is the ``alignment effect", the
tendency of galactic extended emission line regions (EELR) to lie along
the radio axis (see \cite{mcc93} for a review).  At least part of the
explanation involves an AGN in which the radio axis, the dusty torus,
and the photoionization cones are all roughly coaxial. This picture is
generally confirmed with polarimetry, which shows that the reflection
nebula is spatially coincident with the EELR.  Jet-induced star
formation might also contribute to the alignment effect (e.g.
\cite{bes96}).

   In 1994 we started a program of spectropolarimetric observations of
powerful radio galaxies, to study unification and the alignment effect,
and to study the geometry and kinematics of the AGN.  We have already
reported on 8 objects: PKS 0116+082, 3C 234, FSC 10214+4724, 3C 265, 3C
277.2, 3C 324, 3C 343.1, and Cygnus A.  (\cite{coh97}, T95,
\cite{goo96,coh96}, \cite{tra98}, O97) In this paper we give the
combined results for the 13 objects for which $z \le 0.3$; i.e.., those
in which H$\alpha$ is available, and discuss the polarimetry,
unification, geometric effects, and the continuum radiation.  In
general we assume that the scattering is due to dust, which undoubtedly
is present, at least near the nucleus, because of the observed
extinction.  The relative importance of electron and dust scattering is
undetermined, since dust can mimic electrons in producing
wavelength-independent scattering in the optical band.

   The organization of this paper is as follows.  In \S2 we describe
our sample of FR II galaxies, \S3 gives some observational details, and
\S4 and \S5 give a general view of the results. Section 6 presents
a model for the continuum radiation for BLRG.  Details of individual
objects are in \S7 and \S8.  In \S9 we discuss the rotations in
$\theta$ (polarization position angle) observed in 3C 227 and 3C 445.
Section 10 contains further discussions of direct vs scattered light,
unification, and radio-optical connections.  We choose $\rm{H_o}
= 65 ~km ~sec^{-1} ~Mpc^{-1}$ and $q_o = 0.05$ for all calculations. We
shall make use of both the full Stokes parameters $(Q,U)$ and the
normalized parameters $(q,u)$; the use of upper and lower case letters
should minimize any confusion.

\section{The Sample}                   

   We have been measuring the polarimetric properties of radio galaxies
with the 10-m W.M. Keck telescopes, and define those in which H$\alpha$
is visible as the ``low-z sample". To this Keck group we add two objects
observed with the 5-m Hale telescope at Palomar Observatory. The total
sample, listed in Table 1, comprises 13 FR II RG with $z \le 0.3$ -- 5
BLRG and 8 NLRG. 3C 234 is often called a BLRG because broad H$\alpha$
is readily seen in F$_\lambda$, but we list it with the NLRG because the
broad line is predominantly composed of scattered light (See \S4.1).

  Column 6 of Table 1 gives an estimate of an upper limit to the
interstellar polarization (ISP) towards each source.  This value is
obtained by estimating the extinction towards each object from the maps
prepared by \cite{bur82}, and using the empirical formula $p_{max} =
0.09 \times E(B-V)$ (\cite{ser75}).  Cygnus A is at low latitude and is
discussed by \cite{tad90}; see also O97. The ISP effects are small in
this case. Possible ISP effects in 3C 33, 3C 195, and FSC 2217+295 are
discussed in \S8 and \S9.  References to previous polarimetric work are
in column 7 of Table 1.

  Our sample is not an approximation to a flux-limited or
volume-limited sample, as most of the objects were chosen because they
were known to be polarized.  3C 105 and 3C 135 were observed because
they are FR II RG that fitted into a slot in the observing schedule.
3C 357 was observed because its HST image (\cite{dek96}) shows an
off-nuclear emission region that we thought might be polarimetrically
interesting.  We shall make use of work by H96 who observed P$\alpha$
in a complete sample of 11 RG (see \S10.2), 3 of which are also in our
sample. This will give our results more generality than they would have
by themselves.

\section{Observations}

  The Keck observations were made with the polarimetry module
(\cite{goo95,coh97}) in the Low-Resolution Imaging Spectroscope
(\cite{oke95}). The polarimeter contains a rotatable half-wave plate and
a calcite beamsplitter.  The Stokes parameters are measured by taking 4
exposures at waveplate position angles (PA) of $0\deg, 45\deg,
22.5\deg, \rm{and} ~67.5\deg$, and combining the spectra with the
standard self-calibration procedure (\cite{coh97}), using the computer
reduction system VISTA.  At each epoch we observed standard polarized
stars and standard null stars as a check on the system and to establish
the reference for PA determinations. The standard stars are
taken from the lists of \cite{sch92} or have been referenced back to
that list by our own measurements with the Keck polarimeter.  The
relative accuracy in PA determinations is about 0.5\deg.  The PA
calibration is discussed in detail by \cite{ogl98}.

   For spectroscopy we used a grating with 300 g/mm, blazed at 5000
\AA.  This gives a dispersion of $\sim2.5$ \AA\ per pixel and a
resolution of $\sim10$ \AA. The wavelength coverage was typically
3900~\AA\ to 8900~\AA.  The slit was $1.0 \arcsec$ wide, except where
noted in Table 2.  We did not observe with an order-blocking filter.
In most cases the spectra are flat or rise to the red, and errors due
to second-order light are negligible. We did observe the flux standards
with and without the order-blocking filter, as most of them are hot
white dwarfs  and the flux calibration would otherwise have been in
error.

   For imaging we replaced the grating by a mirror and inserted a
B-band filter into the beam.  The scale is $0.215\arcsec$ per pixel.
The imaging polarimetry procedure is the same as for spectroscopy, with
the simplification that there is no spectral extraction.  When
necessary for either spectroscopy or imaging we improved the S/N by
binning the fluxes rather than by averaging the Stokes parameters
themselves, which can lead to bias (\cite{cla83}). We typically took
15-minute exposures so that a full 4-exposure set took an hour plus
readout time.

   The field of view of the Keck polarimeter is about $32\arcsec$. This
limits the useful length of the slit for spectroscopy, and for all
observations it can make sky subtraction difficult, as some galaxies
nearly fill the field.  In such a case the apparent polarization $p$
may be higher than the value a larger field would have given; this is
analogous to increasing $p$ for a nuclear source by narrowing the slit
and allowing less diluting starlight into the the measurement.

   At Palomar we used the double spectrograph (\cite{oke82}) 
to do spectropolarimetry, but not imaging polarimetry, on the 5-m
Hale telescope. The polarimeter is similar to that used at Keck
(\cite{goo91}).  In the red camera the grating has 158 g/mm and
dispersion approximately 6 \AA\ per pixel, and in the blue camera the
grating has 316 g/mm giving approximately 2.1 \AA\ per pixel.
Calibration and data reductions were done in a fashion similar to those
used at Keck.

  Table 2 gives the log of all the observations. The imaging
observations on 1995 December 16 were compromised by poor seeing that
ranged up to $1.8\arcsec$. We reobserved 3C 195 and 3C 234 on 1996
April 16 and only the later data are included in this paper. Four RG,
PKS 0116+082, 3C 109, 3C 227 and FSC 2217+259, were observed only in
December 1995, and those data are included here except for PKS
0116+082, which was presented in Cohen et al. (1997).  The high and
variable seeing compromises the angular resolution, but does not
further affect the polarization observations.

\section{Results: Spectropolarimetry}

  We display the combined spectropolarimetry results in Figures 1 and
2, where the objects are arrayed, roughly, in order of decreasing
visibility of the broad H$\alpha$ line in F$_\lambda$.  In Figure 1 the
left panels show the total flux F$_\lambda$ and the right panels show
the ``polarized flux" defined as $p\times\rm{F}_\lambda$.  This
quantity is sometimes called the ``Stokes Flux", and in simple cases it
is proportional to the scattered flux.  The spectra of F$_\lambda$ and
$p \times \rm{F}_\lambda$ are shown on a logarithmic scale.  Note
that corresponding plots in Figure 1 have the same logarithmic range,
so that the relative enhancement of H$\alpha$ can be more readily
assessed.  Figure 2a shows the fractional linear polarization $p$ and
Figure 2b shows the polarization position angle $\theta$.  When the S/N
is high ($>5$) we can estimate $p$ as $\sqrt{(q^2+u^2-\sigma^2)}$ where
$q$ and $u$ are the normalized Stokes parameters and $\sigma$ is the
error in $q$ or $u$ calculated from photon statistics and the normal
propagation of errors (\cite{sim85}). In most cases, however, S/N is
not that high, especially at the blue end of the spectrum. Instead of
attempting to debias the square-root, we prefer to rotate $q$ and $u$
by $-2\Theta$, where $\Theta$ is a smooth approximation to the measured
$\theta$ (\cite{coh97}).  The resulting $q^\prime$ is a useful estimate
of $p$.  Hereafter we use the symbol $p$ for all estimates of the
polarization.

   In Figures 1 and 2 we show spectra for 3C 109, 3C 234 and Cyg A that
have been published before (\cite{goo92}, hereafter G92; T95, O97) and
are repeated here to give a more complete impression of the range of
possible behaviors.  3C 105 had no measurable polarization and spectra
for it are shown in \S8.3.  The polarization of 3C 357 is particularly
noisy and it is omitted from Figures 1 and 2 but is shown in Figure 3
and discussed in \S8.7.  Generally the spectra are for a region
centered on the nucleus, except for 3C 33 where the extraction window
was centered on a strongly-polarized region approximately 3\arcsec~
north-east of the nucleus, and 3C 357 where the window is approximately
1.5\arcsec~ west of the nucleus. The length of the window is given in
Table 3, which gives some results from the polarimetry.  The averaging
interval, 5400 -- 5600 \AA, contains no strong emission lines, so the
values of $p$ and $\theta$ represent, approximately, the V-band
continuum polarization.  However, we show in \S5.2 that many of the
NLRG contain reflection nebulae, in which $\theta$ varies with
position. In these objects the measured $\theta$ may depend on the
location of the slit and the extraction window.

   For most objects an elliptical galaxy template was fit to the
absorption features and subtracted from F$_\lambda$ to get the
``corrected flux" F$_{\lambda,c}$. The corrected polarization $p_c =
p/(1-f_g)$ is in Table 3.  We used NGC 6702 as the template galaxy
except for 3C 234 and Cyg A where NGC 821 was used.  Two BLRG, 3C 382
and 3C 445, have so little starlight that we were unable to determine
an accurate galaxy component other than $f_g < 0.1$. We were unable to
fit 3C 109 properly, but the absorption lines are weak and the galaxy
fraction must be small.  Figure 3 shows curves of F$_{\rm\lambda,c}$
and $p_{\rm c}$ for 7 objects; those for  3C 234 and Cygnus A have
already been published (T95, O97).

  At this point we discuss the main features of Figures 1-3.  Details
will be presented in later sections.

\subsection{Broad vs Narrow Line Objects and 3C 234} \label{41}

   It has been customary to define a broad-line radio galaxy merely as
one that has prominent broad lines, e.g. the Balmer series, that are
substantially broader than the forbidden lines e.g. [OIII].  This
simple definition, however, takes no account of equivalent widths, 
and is subject to the vagaries of observation and
signal-to-noise ratio.  Further, it does not allow for differing
geometric or physical circumstances. In Figure 1a, 3C~234 shows
prominent broad H$\alpha$ and easily seen broad H$\beta$.  These lines
were noted long ago and thus 3C 234 is usually called a BLRG
(\cite{gra78}). Note, however, that the broad lines are noticeably
weaker than those in the top 5 galaxies in Figure 1a, and 3C 234 has
also been called an NLRG.  \cite{ant90} argue by analogy with the Type
2 Seyfert NGC 1068 that the broad lines are polarized like the
continuum and are predominantly due to scattered light, and that 3C 234
should be called a Type 2 object (see also T95).  We agree that the
broad lines are mainly due to scattering, because broad H$\alpha$ and
H$\beta$ are more prominent in $p\times\rm F_\lambda$ than in
F$_\lambda$. Thus we discuss 3C 234 with the NLRG.  However, it is
intermediate between Type 1 and Type 2 objects, and has other
differences from the BLRG and NLRG in addition to the intermediate
equivalent width of the broad lines. We discuss this in \S8.5.

\subsection{Continuum Radiation} \label{42}

   In Figure 1a the BLRG have been arrayed in order of decreasing
visibility of broad H$\alpha$, but it can be seen that the order is
also one of increasing redness and $p$. 3C 382 has a blue quasar
spectrum and is weakly polarized; FSC 2217+259 is a red IRAS galaxy and
is highly polarized.  The correlations are shown in Figure 7, along
with calculations made with a simple model which is presented in \S6.
Briefly, the model assumes that the BLRG continuum (after starlight
subtraction) is a combination of two components, a nearly unpolarized
$direct$ ray that has a view of the nuclear continuum source and BLR
but has extinction and reddening due to dust, and a highly polarized
$scattered$ ray that has less reddening.  The net polarization
increases as the ratio of scattered to direct radiation increases, so
the direct ray bluens the polarization and reddens the total flux.
This model is similar to the one used by \cite{wil92} for IRAS
13349+2438.  Application to individual objects is given in \S7.

   The continuum model cannot be used for the NLRG because they have an
additional important component of radiation.  Most of the NLRG, along
with many Seyfert 2 galaxies, have broad lines that are more highly
polarized than the neighboring continuum, even after the subtraction of
starlight (Figure~3b). This led Tran (1995) to consider two featureless
continuum components; the first, FC1, is due to scattering and is
polarized, while the other, FC2, is assumed to be unpolarized.  FC2
differentially dilutes the continuum and H$\alpha$; this leaves $p{\rm
_c(H\alpha)} > p{\rm _c(cont)}$.  The relative levels of FC1 and FC2
can be estimated by requiring the broad H$\alpha$ emission line to have
the same intrinsic polarization as the neighboring FC1.  This is useful
if it can plausibly be maintained that the AGN light (continuum and
broad lines) is all polarized at the same level, by scattering.  We
think that this is a conservative assumption, as it is unlikely that
the broad lines could be intrinsically more polarized than the nuclear
continuum. They surely come from a larger region (\cite{tra95a}), and
are more likely to be intrinsically polarized less than the continuum,
as in 3C 382 (see also \cite{martel96,ogl99}).

   In 3C 234 the decomposition into FC1 and FC2 (T95) showed that the
continuum may consist of three components in addition to unpolarized
starlight: Balmer nebular continuum radiation from the NLR, nuclear
light in which the continuum and broad lines are polarized the same
(FC1), and an FC2 whose origin is unclear. The reliability of this
model was enhanced when it turned out that broad H$\beta$ had the same
$p$ as its neighboring FC1, even though H$\beta$ was not used in the
construction.  Cygnus A yields a similar result (O97), and in this case
FC2 can be interpreted as diluting radiation from hot young stars
(\cite{goo89}; see also \cite{hec95,sto98}).

\subsection{$p$ and $\theta$}  \label{43}

   Nearly every object has $p$ rising to the blue (Figure 2a). For most
objects this is mainly due to the reduction of dilution of $p$ by
unpolarized light from an old stellar population.  (This point has been
made numerous times; see e.g. T95, and \cite{cim95}).  It is generally
said that most low-z RG have low $p$ whereas most high-z RG have high
$p$, and that this is due to the different rest-wavelength ranges
covered at low and high z.  Figure 2 shows this wavelength dependence:
at 5500 \AA\ all but 3C 109, FSC 2217+259 and 3C234 have $p < 3\%$,
whereas below 4000 \AA\ all but 3C 382, 3C 445 and 3C 135 have $p >
3\%$.  However, Figure 3b shows that $p_{\rm c}$ still rises to the
blue in most objects. As remarked above, this could be caused by
another diluting component; e.g.  a reddened direct ray or FC2.

   The rise of $p_{\rm c}$ to the blue, especially in the UV, means
that the polarized flux $p\times\rm F_\lambda$ has a steeper spectrum
than the galaxy-corrected flux F$_{\rm\lambda,c}$.  In $p\times \rm
F_\lambda$ 3C 382 and 3C 321 both have an index $\alpha \approx -2.0$
(F$_\nu \sim \nu^{-\alpha}$), at least blueward of H$\alpha$. This is
part of the evidence that 3C 321, an NLRG, has a quasar-like central
continuum source. See \S8.6.

   The position angle $\theta$ (Figure 2b) is remarkably constant in
many of the objects. In FSC 2217+259, for example, a linear fit to
$\theta$ gives a slope $0.0004\deg \pm 0.0003\deg/$\AA~which is at most
$2\deg$ across the spectrum. In 3C 234 a similar fit gives a change
$\Delta\theta = -2.5\deg \pm 4.4\deg$ across the spectrum.  In these
objects there can be little polarization by intervening dust
extinction, either in our galaxy or the host galaxy, because $p$ itself
changes strongly with wavelength and the observed $\theta$ would rotate 
noticeably, unless there were a fortuitous alignment of position
angles.  A few objects do have a significant rotation in the
position angle of the continuum. The two objects with double nuclei
have large rotations; 3C 321 rotates $\sim28\deg$ between 3550~\AA ~and
7000~\AA, and Cygnus A has $\Delta\theta = 19\deg \pm 4\deg$.  The
Cygnus A rotation is explained with three components of continuum
radiation, with different $p$'s and $\theta$'s, that combine to give
the observed values (O97).

\subsection{Polarization of the Broad Lines} \label{44}

   In all the polarized NLRG except 3C 357, $p$ rises in broad
H$\alpha$ (Figure 2a). This is due to differential dilution by
starlight and other components.  It is this increase in $p$ that makes
broad H$\alpha$ more prominent in the polarized flux $p \times
F_\lambda$ than in $F_\lambda$ (Figure 1).  3C 357W is the only
polarized object that does not show broad lines.  However, this object
is faint and the telluric A band falls on the red wing of H$\alpha$, so
this result is inconclusive (see \S8.7).  3C 382 is different
from the other BLRG as it has a quasar spectrum, with a weakly
polarized continuum and little or no starlight or narrow lines. Broad
H$\alpha$ and H$\beta$ are weak in $p \times F_\lambda$, and are nearly
unpolarized.  Weak polarization of the broad lines relative to the
continuum is typical of quasars (\cite{ogl99}).

  The sequence in Figure 1b shows that broad H$\alpha$ is similar in
polarized flux from 3C 227 to at least 3C 33.  In these cases the
polarized flux is giving us a look deep into the interior of the
object, by scattering. We also see the interior of the BLRG directly,
but with extinction (see \S6). This may be true also for 3C 234 (T95).
In 3C 234, and in some of the BLRG, the LOS is probably near the edge
of the obscuring torus. We suspect that the torus does not have a sharp
boundary, and that there is a range of inclinations where the interior
can be seen through dust.

\subsection{Narrow emission lines}  \label{45}

   In general the narrow lines are only weakly polarized. This is
readily seen in Figure 2a, where many objects show strong reductions in
$p$ at the locations of the narrow lines. This is most striking in 3C
234 where more than a dozen narrow lines show the effect strongly.
Narrow H$\alpha$ shows up prominently as a reduction in $p$ in 3C
234, 3C 33NE, and 3C 195  because these are the objects that are
dominated by narrow lines in total flux but by broad lines in polarized
flux.

   All the objects except 3C 109, 3C 234, and 3C 382 show narrow
emission lines in $p \times F_\lambda$ (Figure 1b), so these lines are
polarized but at a low level. The possible polarization of the [O~III]
lines in 3C 109 is discussed in G92.  Weakly polarized
[O~III]$\lambda\lambda4959,5007$ is seen in most objects, several
objects show polarized narrow H$\alpha$, and Cyg A shows several other
species. [O~II]$\lambda3727$ is not reliably identified in any of the
$p \times F_\lambda$ spectra, and is essentially unpolarized.  The
origin of [O~II] must be at or outside the scattering region, whereas
the other narrow lines arise from deeper locations and are partly seen
by scattering.  The possible anisotropy of [O~II] and [O~III] and its
interpretation in terms of unified AGN models has been discussed by
e.g. \cite{jac90,hes93,bak97}, and \cite{dis97}.

\section{Results: Imaging Polarimetry} \label{5}

   We took polarimetric images of 11 objects (Table 2).  Earlier
observations of this type are referenced in Table 1. Our current work
shows that the polarization images fall into two main classes, those
with compact and those with extended polarized regions.  There is a
strong correlation between this image class and the spectral
classification: the BLRG are compact and the NLRG are extended.

\subsection{BLRG} \label{51}

  The compact systems are shown in Figure 4.  The innermost contour is
50\% of the peak for every object, and except for FSC 2217+259 the next
contour is 5\% of the peak (see the caption).  All the objects except
FSC 2217+259 have $\rm{FWHM} \approx 1\arcsec$, consistent with
seeing.  3C 109 and 3C 227 have been called ``N" galaxies
(\cite{spi85}), indicating that they contain a brilliant point-like
nucleus.  Other N galaxies in our sample include 3C 135 and 3C 234,
shown in Figures 5 and 6, and 3C 445, for which we have no imaging
data.  3C 382 has a strong nucleus although apparently it was not
called an N system.  In \S7.4 we give evidence that it was
exceptionally bright at our epoch.

   3C 382, 3C 109, and 3C 227 are closely similar. The galaxies are
dominated by a bright point source at the nucleus, and the polarization
patterns are simple.  The vectors are roughly parallel and decrease
radially by a small amount -- a factor of 0.8 in 3C 109 and 3C 382, and
0.6 in 3C 227. These patterns can be explained with a polarized point
source broadened by seeing and diluted by galaxian starlight and by
spatial averaging.  There is no need to invoke scattering by an
extended distribution of dust or electrons to explain the observations.
FSC 2217+259 is highly polarized and consequently has high S/N. The
vectors are accurately parallel and decrease radially by a factor of
two or three. It is similar to the other BLRG and the polarization
pattern could again be due to seeing plus dilution, although the
angular extent is greater than for the others and scattering in an
extended medium may also play a role.  This object is very red and we
are seeing the nucleus through dust, as discussed in \S7.4.

   In 3C 382 $p({\rm H\alpha})<p({\rm cont})$  and the scattering region
is largely interior to the BLR. It may not be coincidental that 3C 382
has the weakest polarization of the BLRG. In 3C 109 and FSC 2217+259
$p({\rm H\alpha})=p({\rm cont})$ and $\theta$ does not change across
H$\alpha$; the main scattering region must be well outside the BLR. In
3C 227 and 3C 445 $\theta$ rotates in H$\alpha$; this is discussed in
\S9.

\subsection{NLRG}   \label{52}

   Images for the NLRG are shown in Figures 5 and 6, except for Cygnus
A, which has been extensively discussed elsewhere (O97). In Figure 5
the top panels show the total flux, the polarization vectors, and the
radio axis. The lower panels show contours of polarized flux
$p\times\rm{F}_\lambda$, with the slit location. Figure 6 shows total
flux and polarization vectors for the two NLRG that have weak
polarization and do not have extended polarization regions.  In 3C 33,
3C 195, 3C 321, and Cyg A (O97)  the polarized regions are in ``fans"
on opposite sides of the nucleus, and the polarization vectors are
roughly azimuthal; i.e. they are tangential to circles around the
center.  This is the sign of a bipolar reflection nebula. A central
source illuminates material in a bicone, and the scattered light is
polarized perpendicular to the plane of scattering.  Although the
perpendiculars to the polarization vectors accurately define a center
when averaged, seeing plus spatial averaging tilt the individual
vectors away from perpendicular.

  In all the objects with polarization fans the polarization is weak at
the nucleus, and has a strong radial gradient. The sign of the gradient
is opposite to that for the BLRG, where $p$ decreases with radius.  The
gradient must be partly due to dilution by unpolarized starlight, which
is more strongly concentrated to the nucleus than is the scattering
material.  However, the effect is also seen in 3C 265 with z=0.811
(\cite{tra98}).  In this case the polarization image was measured in
the rest-frame UV, and dilution by an old stellar population would have
had little importance.  Seeing and spatial averaging also affect the
pattern, but they generally act to reduce the gradient.

   Three of the four objects in Figure 5 and also Cyg A (O97); i.e. all
the strongly polarized NLRG except for 3C 234, have separated peaks in
$p\times\rm{F}_\lambda$. Since starlight is unpolarized it has no
effect on $p\times\rm{F}_\lambda$ and we must look elsewhere for the
cause.  We suggest it is due to the presumed geometry, an illuminated
cone whose apex is shielded by an opaque torus. Near the apex the
resolution element is larger than the cone and the vector sum of the
polarizations is reduced because they have a variety of angles, all
perpendicular to the radius. The central decrease could also be caused
by obscuration of the polarized light.

  3C  234 is special. The polarization pattern is extended but does not
show fans, and $p$ has a steady EW gradient with a peak east of the
flux peak.  $p\times\rm{F}_\lambda$ has a maximum at the center.  This
pattern is different from both the BLRG and the NLRG.  The vectors
show a tendency for circular symmetry, although not as strongly as the
other objects, and the pattern may be due to scattering of nuclear
light.  This is discussed further in \S8.5.

\section{BLRG Spectral Index, Polarization and Reddening} \label{6}

  In this section we first discuss the spectral index and Balmer
decrement for the BLRG in terms of one reddening screen, and then
present a two-component model that also attempts to explain the
polarization.  In \S7 the results are applied to the individual
sources.

\subsection{One-Component Model - Absorption Only} \label{61}

   Assume that there is a dust screen in front of the BLRG which
produces the same extinction in the continuum and the lines. Then we
have two measures of reddening: the ratio of broad H$\alpha$ to broad
H$\beta$, and the ratio of the continuum fluxes at H$\alpha$  and
H$\beta$.  The data plotted in Figure 7a show that these quantities are
correlated.  They have been corrected for Galactic reddening, and 3C
227, 3C 234, and FSC 2217+259 have been corrected for an old stellar
population (the other objects have $f_g <0.1$).  The continuum spectral
index $\alpha$ between H$\alpha$ and H$\beta$ is shown in the top
scale.  In Fig 7 the order of the objects (left to right) is the same
as the vertical order in Figure 1 except that Galactic dereddening has
moved 3C 109 to the left of 3C 445.  Baker (1997, Fig.  16) shows a
similar diagram for radio-loud quasars. The range of Balmer decrement
in Figure 7a is about the same as in Baker, but FSC 2217+259 has a
considerably redder spectrum than any of her quasars.

   Four of the five BLRG (excluding 3C 109) form a sequence that is
parametric in reddening. We model this relationship by assuming that the
intrinsic Balmer decrement of the broad-line clouds is 3.0 (Case B).
The reddening line for this model depends on the intrinsic spectral
index $\alpha_{\rm o}$ of the continuum source, and in Figure 7a we
plot solid lines for ${\rm\alpha_o=-0.5,~0.0,~0.5,~and~1.0}$. The great
majority of quasar indices fall within this range (\cite{fra91,bak95}).
The dotted lines show extinction ${\rm A_V}$ in magnitudes, calculated
with the \cite{car89} extinction law with $R_V = 3.1$.

   The model can be adapted to all the BLRG except 3C 109.  However,
radio loud quasars (excluding CSS sources) appear to have dereddened
intrinsic indices close to 0.5 (\cite{bak95}), and FSC 2217+259 lies
rather far from that value. It may be that it has an exceptional index,
but it may also be that the model is too simple in assuming that the
extinction is the same for both the continuum and broad lines.  If the
extinction to the BLR were more than to the continuum source, then FSC
2217+259 could be accommodated with an index near 0.5.  Indeed, the extra
extinction could be associated with the BLR itself.  For a
discussion of dust in the BLR see e.g. \cite{goo95a}.  Another
alternative is that the intrinsic Balmer decrement is substantially
higher than 3.0 \cite{kwa81}.

   3C 109 is well outside the range of the model. It is exceptional,
either in the intrinsic Balmer decrement or the intrinsic index (or
both).  Alternatively, the model may not apply to 3C 109, at least not
in this simple form.  3C 234 is shown for comparison only; the model
cannot be used for the NLRG because the scattered light arises far from 
the nucleus and cannot be presumed to be behind the nuclear screen.
Also, the NLRG show an important radiation component, FC2, that is
not accounted for here.

\subsection{Two-Component Model} \label{62}

  Figure 7b shows the galaxy-corrected continuum polarization at 5500
\AA ~(rest) vs ${\rm F_{\lambda,c}(H\alpha) / F_{\lambda,c}(H\beta)}$
for the five BLRG and 3C 234.  There is a trend for the polarization to
increase with reddening. \cite{rud83} made a similar plot for 13 BLRG;
their sample includes all of our objects except FSC 2217+259.  They
explained the correlation between $p$ and the Balmer decrement as
either dichroic absorption or scattering with extinction. The hidden
nucleus model for NLRG leads us to consider a more elaborate model to
explain the correlation. It consists of two components: an unpolarized
direct ray, and a ray from the nucleus that is scattered and highly
polarized.  The components both have extinction, but the direct ray has
more. This is consistent with the presumed BLRG geometry in which we
are near the edge of the cone. The direct ray must penetrate more dust
than the scattered ray that passes closer to the axis. The effect of
the model is to redden the continuum, increase the Balmer decrement,
and bluen the polarization by differential dilution.

    The model is similar to that used by Wills et al. (1992; see also
\cite{bro98}), except we assume that the extinction of the scattered
component arises from dust inside the scattering volume itself.  This has
the advantage that the scattering and extinction are coupled, and there is
one less free parameter.  The direct ray can be expressed as ${\rm F_d =
F_o e^{-\tau_d}}$, where ${\rm F_o}$ is the unobstructed flux from the
nucleus, and ${\rm\tau_d}$ is the optical depth.  To properly describe
the scattered ray we should specify a geometry and integrate through the
scattering volume.  Such calculations have been done by \cite{man96}
for a bicone uniformly filled with dust, but we merely make the simple
approximation used by Fosbury et al. (1999). We assume that the
scattered light can be expressed (approximately) as ${\rm F_s = F_o k_g
Q_s e^{-\tau_{ext}}}$, where ${\rm k_g}$ is a geometric factor and ${\rm
Q_s}$ is a scattering efficiency that depends on the size distribution of
the dust particles and on their composition.  Each ray passes through
a dusty region before and after it  is scattered, and the factor ${\rm
e^{-\tau_{ext}}}$ accounts for the extinction in the scattering volume.
The optical depth ${\rm\tau_{ext}}$ is proportional to an extinction
efficiency ${\rm Q_{ext}}$ that is roughly proportional to the scattering
efficiency ${\rm Q_s}$, provided the wavelength is not near the graphite
resonance at 2200 \AA.  The net result of these assumptions is that the
scattered ray can be expressed as

    $${\rm F_s = F_o {\rm k_s} \tau_{ext} e^{-\tau_{ext}}} \eqno{(1)} $$  

   The function $\tau_{ext} e^{-\tau_{ext}}$ has a maximum at ${\rm
\tau_{ext}=1}$, and most of the contributions to F$_{\rm s}$ come from
regions near ${\rm \tau_{ext} \approx 1}$.  For simplicity we now set
${\rm \tau_{ext}=1}$ and Eq (1) becomes

           $${\rm F_s = 0.37 {\rm k_s} F_o }               \eqno{(2)} $$

\noindent where ${\rm k_s}$ is independent of wavelength except for
geometric factors associated with the change of the $\tau_{ext} = 1$
surface with wavelength. ${\rm k_s}$ is an adjustable parameter; we
expect ${\rm k_s<<1}$ and in the models shown below we use
${\rm~k_s}=0.05$.

   The scattered flux ${\rm F_s}$ can be highly polarized; the curves
by \cite{man96} show that the intrinsic polarization on ${\rm F_s}$,
$p{\rm _s}$, can be as high as 50\%.  The apparent polarization
$p_{\rm~c}$ is

      $$p_{\rm c} = p_{\rm s}{\rm F_s/(F_s + F_d)} \eqno{(3)}$$ 

\noindent Equation (2) states that the scattered flux is proportional
to the incident flux; what we have done is to crudely justify the
common assumption that the intrinsic polarization is independent of
wavelength. The calculations by Manzini \&\ di Serego Alighieri,
however, show that $p_{\rm s}$ generally rises to the red.

   The model can be parameterized in various ways. In Figure 7b we show
lines of constant ${\rm \alpha_o}$ (solid), and horizontal lines of
constant extinction for the direct ray, ${\rm A_V(d)}$ (dotted),
together with the corrected data.  In this example the fixed parameters
are ${\rm k_s=0.05}$ and $p_{\rm s} = 0.3$.  The model also predicts the
Balmer decrement and this is shown in Figure 7c, which is analogous to
Figure 7a.  The Balmer decrement and the index are limited because
${\rm \tau_{ext} = 1}$, and as ${\rm A_V(d)}$ increases beyond 3 mag
the scattered ray becomes dominant.

   3C 109 and FSC 2217+259 are outside the range of the model in Figure
7c, and they are at the extreme values of ${\rm \alpha_o}$ in Figure
7b.  The BLRG of course must have a range in their intrinsic properties,
but it is difficult to fit 3C 109 and FSC 2217+259 into
the scheme described here.  Enlarging the model by adding excess
extinction to or in the BLR is one way to accommodate them. Another is
by postulating the existence of an additional blue continuum component.
In both cases it must be done in a way that preserves the near equality
of $p_{\rm c}$ for H$\alpha$ and the continuum; i.e. the equivalent
widths of the broad lines must be similar for all the rays.  In
studying the reddening of radio-loud quasars, \cite{bak97} found
analogous discrepancies.  She suggests that excess reddening to the BLR
could be the explanation, but that other factors could also be
responsible.  Optical depth effects that increase the intrinsic Balmer
decrement could be part of the answer, as could dichroic dust
extinction.  See \S7.1.

   The point to this exercise is to show that the simple two-component
model is able to explain qualitatively the observed spectra and polarimetry
of the BLRG.  Further insight could be obtained with detailed
calculations and a specific geometric model, and by examining the full
spectrum rather than just the two-point spectral index.

\section{Individual Objects: BLRG}   \label{7} 

\subsection{3C 109} \label{71}

  The Palomar spectropolarimetry of 3C 109 is discussed by G92 who
interpreted the polarization as due to transmission through aligned
dust in the host galaxy. In view of the success of the two-component
model, we now think that the origin of the polarization in 3C 109
should be reinvestigated.  3C 109 shows a Serkowski-type curve of
$p(\lambda)$ (Figure 2a), but the data are noisy below 5000 \AA. The
appropriate way to discriminate between scattering and dichroic dust
extinction is by extending the spectropolarimetry into the UV.

   In Figures 7a and 7c 3C 109 stands apart from the other BLRG.  We
can describe this by saying that the Balmer decrement is higher than
expected from the continuum slope. This could result from an
intrinsically high Balmer decrement, or from extra reddening to the
BLR. Alternatively, we could say that the continuum slope is bluer than
expected from the Balmer decrement, and this could result from an extra
blue continuum component that is not in the model. Evidently there is a
number of possibilities for explaining the discrepancy; and, in
addition, there remains the possibility that dichroic extinction plays
an role.

\subsection{3C 227}  \label{72}

   \cite{pri93} have done extensive modeling of the EELR in 3C 227.
They show that the velocity field can be modeled (roughly) with a
bicone of opening angle $120\deg$, tilted $40\deg$ to the plane of the
sky, and rotating about an axis offset $20\deg$ from the cone axis.
The projected bicone axis is at ${\rm PA=40\deg}$ and the projected
rotation axis is at ${\rm PA=60\deg}$.  We look inside a cone but near
the edge.  Their analysis of the emission lines shows that there is a
photon deficit, and they suggest that 3C 227 harbors a quasar. Although
the broad lines are prominent the quasar is ``hidden" because most of
the ionizing radiation is not directly visible.

   The optical continuum radiation is polarized at $41\deg$, which is
parallel to the bicone axis. This is perpendicular to the orientation
expected from scattering in the cone, but is consistent with the
orientation expected from an optically thin disk coaxial with the
cone.  The radio axis is at $85\deg$, which does not seem to be closely
related to any of the above angles. The broad lines are polarized at
$\sim 20\deg$ (see \S9) which is $20\deg$ (in projection) from the cone
and continuum radiation axes.  $\theta$ shows notable rotations in
the broad Balmer lines; we discuss them in \S9. The rotation in
H$\alpha$ has already been noted by \cite{cor98}.

   The spectrum of 3C 227 can be qualitatively understood with the
one-component model in \S6.  From Figure 7a, the Balmer decrement
implies 1.1 magnitudes of extinction, and the color is then explained
with ${\rm \alpha_o = 0.5}$, the typical value for radio-loud quasars.
To also explain the polarization, especially the fact that $p_{\rm c}$
rises to the blue, we need to use the two-component model. Figure 7b
shows that ${\rm\alpha_o \approx -0.2~and~A_V(d) \approx 1.7}$. These
particular values can be modified by varying the other parameters in the
model, but ${\rm\alpha_o}$ remains somewhat low. With ${\rm A_V(d)=1.7
mag}$, $p_{\rm o} = 0.3$ and ${\rm k_s=0.05}$ the calculated apparent
polarization rises from 1.7\% at 7000 \AA\ to 4.6\% at 4000 \AA. This
rise is steeper than shown in Figure 3b, but the result is qualitatively
correct.  To use the model to match the polarization, ${\rm A_V}$ would
have to be reduced to near 1 mag, and the product ${\rm k_s} p_{\rm o}$
would have to be doubled.

   The point-source polarization image in Figure 4 supports the
scenario of \cite{pri93} described above, in which we are looking into
the cone but near the edge. Their photon deficit is `` ... about a
factor 2 for regions closer than 50 kpc, and a factor 13 for those
located at much larger distances." Our estimate of 1 -- 2 mag for the
V-band extinction to the nucleus is in rough agreement with their
estimate of the photon deficit. They conclude that the intrinsic 
luminosity of 3C 227 could put it in the quasar class.

\subsection{3C 382} \label{73}

\subsubsection{Aperture Photometry}  \label{731}

   Our imaging observations of 3C 382 were saturated in three of the
four exposures in the $extraordinary$ beam of the polarimeter, even
though we reduced the exposure substantially from the expected value.
They were not saturated in the $ordinary$ beam because the throughputs
of the two sides differ by about 5\%. And, as seen in Figure 4, the
image shows the hexagonal diffraction pattern due to the secondary
support and the mirror shape. Clearly, 3C 382 contained an unusually
bright point-like nucleus during our observations.  That night we also
observed the standard stars PG 1525-071A and B (\cite{lan92}) and used
them as references. The 2 stars were self-consistent to better than
0.01 mag. We measured fluxes within a $2.1\arcsec$ aperture. The stars
were observed about 4 hours before 3C 382. Two other radio galaxies
were observed between the stars and 3C 382, and each had 4 exposures
that show variations of 0.05 mag.  The night, however, was not
photometric.  With this caveat we find $\rm{B} \approx 14.5$ and
$\rm{M_B \approx -23.0}$ (allowing 0.3 mag for Galactic extinction),
for the night of 16 April 1996 (UT).

\subsubsection{Spectroscopy} \label{732}

   3C 382 shows double peaks in H$\alpha$; this has been interpreted in
terms of a rotating disk (\cite{era94a}).  We made the
spectropolarimetric observations on 19 May 1996 (UT).  Flux
calibrations were performed with HZ 44, but the night was not
photometric.  Figure 1 shows the observed spectrum.  Within our sample,
3C 382 is exceptional in several respects.  F$_\lambda$ rises strongly
to the blue, the narrow [O~III] lines are very weak, and the broad
lines are almost as wide (at least 26,000 km/sec FWZI) as the widest
found in quasars (\cite{kin91}).

   We can compare our flux values with those of \cite{cor98} who
observed the H$\alpha$ region of 3C 382 with a $1\arcsec$ slit on 6/7
June 1995. Their value for the continuum flux at 6500 \AA\ (observed)
is about 7 mJy (Figure 1 in Corbett et al.). Our value is 30 mJy. Corbett's
value for the total flux at the peak of H$\alpha$ (actually, the valley
between the two peaks) is about 23 mJy; ours is 92 mJy. This comparison
is difficult because the shape of H$\alpha$ is known to be variable,
and also the result will depend on seeing, atmospheric transparency,
and on the relative amounts of nuclear and extended light that are in
the extracted spectrum.   In spite of these concerns, the comparison
does suggest that in May 1996 both the continuum and H$\alpha$ were
stronger by a factor of about four than they were in June 1995. We can
also make a comparison at H$\beta$ by using the results of
\cite{ost76}.  Osterbrock et al. measure 114 \AA\ for the total
equivalent width, EW, of H$\beta$  and $2.6 \times 10^{-13}$~erg cm$^{-2}$
sec$^{-1}$ for the total flux in broad H$\beta$.  We measure EW=89
\AA\ and F$=27.2 \times 10^{-13}$~erg cm$^{-2}$ sec$^{-1}$.
The continuum and broad lines appear to have been exceptionally high in
May 1996, about 4 times higher than their values a year earlier and
about 10 times higher than their values in 1974.

   Our observed flux at 5530 \AA\ is 26.3 mJy, which gives ${\rm V
\approx 12.9}$. 3C 382 may even have been the brightest quasar in the
sky at that epoch.  In NED 3C 382 has flux values ranging from 1.6 to
7.3 mJy at 5530 \AA, so we saw 3C 382 in April and May 1996 at a
historic high.  Correcting to rest wavelength and using 0.3 mag for the
Galactic extinction gives $\rm{M_B \sim -25.0}$. In its brighter phases
3C 382 is a highly-luminous quasar.

\subsubsection{Polarimetry} \label{733}

  In the B-band polarization image (Figure 4) the outer vectors are
somewhat circumferential and probably are artifacts due to the strong
flux gradient and variable seeing. The polarization values of the
central pixels are unreliable due to saturation.

   In Figure 2a $p$ rises to the blue, as is common in quasars
(\cite{ogl99}).  This is intrinsic and not due to decreasing dilution
by galactic starlight, as the starlight fraction is $<10\%$ (Table 3).
The broad lines are polarized substantially less than the continuum,
consistent with the hypothesis that we see the continuum source
and the BLR more-or-less directly; the former is polarized by local
scattering, with possibly a synchrotron contribution, and the latter
is weakly polarized by scattering in the BLR itself or farther out.
In our work the BLRG $\theta$'s generally show little correlation with
the radio axis (see \S10), but 3C 382 is the one closest to parallel,
with $\Delta\theta = 11\deg$. This might represent a case of scattering
in a disk (Antonucci 1984).

   The polarization of broad H$\alpha$ has been studied by
\cite{cor98}. Although the flux of both the continuum and H$\alpha$
rose substantially between their measurements and ours, the
polarizations are similar.  We measure $p=0.85\%~\rm{and}~0.65\%$ in
the continuum at 6500 \AA~and 7500~\AA, respectively, and $\sim 0.4\%$
at the bottom of the dip in $p$ caused by dilution with H$\alpha$.
Corbett et al. (their Figure 1) show $1.05\%, 0.85\%, \rm{and} \sim
0.3\%$. We have $\theta = 63\deg \pm 0.4\deg$ with no change at
H$\alpha$, while Corbett et al.  have $60.2\deg \pm 1.1\deg$ with
perhaps a small change in H$\alpha$.  (Corbett et al. decompose the
continuum and H$\alpha$ but here we give values appropriate to the
total flux.)  The $\theta$'s are the same but our values for the
continuum $p$ are lower by a small amount. This difference may be
significant, but the major result is that the polarization changed
little while the flux, both continuum and H$\alpha$, increased
substantially.  This suggests that the geometry did not change and
shows that there is no appreciable starlight even at the fainter
levels, for otherwise the polarization would have increased with the
decreasing dilution.

   Figure 7 shows that the Balmer decrement and spectral index are
consistent with little or no dust along the LOS to the continuum source
and the BLR. The polarized flux (Figure 1b) has $\alpha = -2.0$, and is
substantially steeper than the total flux. $p\times \rm F_\lambda$ is
not an achromatic version of F$_\lambda$ but has been bluened.

\subsection{FSC 2217+259}  \label{74}

   The image of FSC 2217+259 in Figure 4 shows a bright star $6\arcsec$
E, and there is another $13\arcsec$ SE. These stars have polarizations
$0.4\% \pm 0.1\%$ at $\theta = 32\deg \pm 1\deg$ and $0.6\% \pm 0.1\%$
at $37\deg \pm 1\deg$, respectively. This might well be due to
ISP, and so the observations could be corrected
for the average, $0.5\%$ at $34\deg$.  The correction is small,
$<1\deg$ in $\theta$, and does not materially change either the image
or the spectra. We do not correct Figures 1 and 2 for the ISP.  Recall
that in \S4.3 we noted that $\theta(\lambda)$ is flat for FSC 2217+259,
and that this implies that a substantial foreground polarizing agent is
unlikely to exist.

   The polarization is high in FSC 2217+259, rising to nearly 9\% at
4000 \AA\ in $p$, and to 13\% in $p_{\rm c}$.  Figure 2a shows that
H$\alpha$ and H$\beta$ are polarized more than the neighboring
continuum, and $p$ dips in narrow H$\alpha$. In this respect FSC 2217+259
is like 3C 234 and other NLRG, but the effect is much weaker and for
later discussion we will assume that in FSC 2217+259 the broad Balmer
lines are polarized the same as the continuum. The galaxy-subtracted
polarization (Figure 3b) is like that of 3C~227; $p_c$ is reduced in
the narrow [O~III] lines and is nearly the same in H$\alpha$ and the
continuum. FSC 2217+259, however, does not have the $\theta$ rotations
in H$\alpha$ that are displayed by 3C 227.  

\subsubsection{Origin of the Polarization} \label{741}

   It is unlikely that the polarization in FSC 2217+259 is due to
extinction by aligned dust grains, for two reasons. First, $p$ has no
sign of the Serkowski spectrum, which is gently curved with a peak,
usually in V-band (\cite{ser75}). Secondly, the high value of $p$, 9\%
at 4000 \AA, is extreme although not unprecedented in stars close to
the Galactic plane.  VI~Cyg 12, at ${\rm b^{II}=0.8\deg}$, has
polarization similar to that of FSC 2217+259, but is much redder.  It
is more likely that the polarization in FSC 2217+259 is due to
scattering.

   Figure 7a shows that the high Balmer decrement can be explained with
3.8 magnitudes of extinction, but this one-component model leaves us
with a very blue intrinsic spectral index, and says nothing about the
high polarization. The two-component model in Figure 7b, with ${\rm k_s =
.05}$ and $p{\rm _c =0.30}$ yields ${\rm A_V\sim 3.1 mag}$ and
${\rm\alpha_o \sim 1.1}$.  With these parameters the calculated
apparent polarization rises from 4.1\% at 7000 \AA\ to 16.4\% at 4000
\AA. As with 3C 227, the calculated polarization is steeper than the
measured curve in Figure 3b. To match the observations the product
${\rm k_s} p_{\rm o}$ would have to be raised by a factor of about 5. This seems
unacceptably high.  The model gives a qualitative sense of the spectrum
and polarization but does not match FSC 2217+259 in detail.

   The $\sim3$ magnitudes of extinction on the direct ray means that if
this object were seen along a dust-free direction, it would have an
absolute magnitude M$_V \sim -23$.  Further, the broad-line ratio
H$\alpha / \rm{H}\beta$ drops from 10.2 to 4.3, closer to Case B, upon
dereddening. All this is consistent with a quasar seen through dust.

\subsection{3C 445} \label{75}

   We do not have a polarization image of 3C 445. The
spectropolarimetry was taken at Palomar and the noise level is higher
than for the Keck data. 3C 445 is like 3C 382 in having very little
starlight; but is different in having strong narrow lines and a redder
continuum.  The polarization rotation in broad H$\alpha$ is discussed
in \S9. The polarization of the H$\alpha$ region has also been recently
discussed by \cite{cor98}. Our values for F$_\nu,~p,~\rm{and} ~\theta$
(July 1994) agree well with theirs.

   3C 445 has $\rm m_B =17.0$ (NED) which, with a Galactic extinction
correction of 0.2 mag, gives ${\rm M_B =-20.4}$. The internal
extinction at V-band, from Figure 7, appears to be between about 1.1
and 2.5 mag.  Assuming a correction to B-band of 0.3 mag, we derive an
intrinsic ${\rm M_B}$ between -21.8 and -23.2 mag. From this analysis,
based on the Balmer decrement and the continuum polarization, we find
that 3C 445 could intrinsically be bright enough to fit into the quasar
category $({\rm M_B}<-23)$.

\section{Individual Objects: NLRG} \label{8} 

\subsection{3C 195} \label{8.1}

   We first discuss 3C 195, the prototypical NLRG showing broad
Balmer lines and spatial symmetry in polarized light.

\subsubsection{Interstellar Polarization} \label{811}

   In Figure 2b $\theta$ rotates about $8\deg$ between 3500 and 7500
\AA.  This might be due to interstellar polarization, as the galactic
latitude of 3C 195 is low and the estimated upper limit on ISP is 1.1\%
(Table 1). We did not measure stars in the field of 3C 195, but two
stars within $3\deg$ have $p \sim 0.6\%$ at $\theta \sim -35\deg$, and
$p\sim0.9\%$ at $\theta \sim -13\deg$. (\cite{mat70}).  Correction for
the average, $p = 0.75\%$ at $\theta = -24\deg$, yields a curve of
$\theta$ that tilts a little in the other direction, indicating perhaps
an over-correction. In addition the [O~III] lines become less prominent
in $\theta$ than they are in Figure 2, and a $\theta$ rotation in broad
H$\alpha$ becomes noticeable. The partial disappearance of the [O~III]
lines is a sign that the correction may be useful, but the appearance
of broad H$\alpha$ in $\theta$ seems like an artifact. Altogether, we
believe that a correction is needed, one rather smaller than the
average shown above.  However, given the notorious gradients of
ISP in the Galaxy, we do not pursue this farther,
since it needs the measurement of stars in proximity to 3C 195. The
corrected value of $\theta$ is probably close to $120\deg$. The
correction would have negligible effect in B band because $p$ is high,
and Figure 5 is unaffected by it, except in the regions where $p$ is
low. The magnitude $p$ (Figure 2) has negligible correction, at all
wavelengths except the [O~III] lines, because the correction vector (in
$q,u$ space) is nearly perpendicular to the measured vector.

\subsubsection{The Polarization Image} \label{812}

   Figure 5c shows the B-Band polarization image. The image shows
oppositely directed ``fans" of polarization with vectors that are
roughly azimuthally directed.  This is the sign of a reflection nebula,
in which a point source illuminates the surrounding material and the
scattered light is polarized perpendicular to the plane of scattering.
The fans are roughly co-spatial in projection with the EELR
(\cite{cim95}). The general interpretation of this pattern is that an
anisotropic beam of radiation illuminates and photoionizes the
surrounding material in two oppositely directed ``cones". The EELR fans
are the projected cones of ionization.  The beam is scattered
by electrons and dust in the EELR and we see a partially polarized
version of the nuclear spectrum. The narrow lines are recombination
radiation that is less polarized than the nuclear light because it
arises outside (or perhaps partly inside) the scattering region.

   Note that the cones are not symmetric. To the NW there appears to be
a non-illuminated region that extends in close to the nucleus.  To the
SE the corresponding region has polarized light, close to the nucleus.
This may reflect different S/N on the two sides, rather than a real
lack of scattered light NW.

   In Figure 5c $p$ is weak in the center (3\%), rises rapidly for
$2\arcsec$ to the NE (15\%) and SW (11\%), and then decreases slowly.
The strong gradient is somewhat surprising, since a simple model,
consisting of optically thin scattering from electrons or dust with
density decreasing linearly with radius, gives $p$ that is independent
of radius.  Figure 5d shows a contour diagram of the polarized flux, $p
\times \rm{F}_\lambda$, whose axis is oriented at PA$\approx 35\deg$,
rather far from the radio axis at $19.5\deg$.  $p\times\rm{F}_\lambda$
has two peaks; the stronger one is displaced about $0.5\arcsec$ SW of
the flux peak, and the other is $2\arcsec$ NE, where the flux contours
are extended and bend N.  The reduction in $p$ and
$p\times\rm{F}_\lambda$ at the center is discussed in \S5.2, where we
suggest that it is due to the cone geometry, with dilution by starlight
also affecting $p$. This reduction in $p$ and $p\times\rm{F}_\lambda$
is common, as it exists in 4 of the 5 highly polarized NLRG.

   A uniformly filled scattering cone will show a polarization at the
apex that is perpendicular to the axis of the system.  In the central
$1.3\arcsec$ of Figure 5c the vectors have
$<\theta>=130.4\deg\pm3.7\deg$, and we assume they are perpendicular to
the axis of the cone, which then is at PA~$\approx 41\deg$. The axis of
$p\times\rm{F}_\lambda$ (35\deg) is close to this cone axis. A line at
${\rm PA=41\deg}$ is within about 2\deg of perpendicular to all the
vectors along its path.  This line is the symmetry axis of the
reflection nebula, because of the perpendicular polarization at the
apex, and also the close perpendicularity along the line which is not
matched elsewhere.

   The axis, and the rotation in $\theta$, can be appreciated in Figure
8, which is a color-coded diagram of $\theta$. The rotation is clear in the
north, where the colors are roughly radial, and a center can
approximately be located where the various sectors come together. This
point is the illumination center. It appears to lie between the peaks
in $p \times \rm F_\lambda$, and is not at the flux peak, but rather is
displaced to the NE by $0.7\arcsec \pm 0.2\arcsec$.  This situation may
be analogous to the displacements in Cygnus A, where the center is
hidden by dust (O97). It would be interesting to compare the radio and
optical positions for 3C 195 to see if there is an offset, but the
optical positional uncertainty is comparable with our measured
displacement (\cite{dis94b}).

   The pattern in Figure 8 is more regular in the N than the S, but
the axis can be traced across the entire object.  In the E and W the
red regions are roughly perpendicular to the radius, but they do not
extend into the central region.  It is likely that the nuclear region
is complex, as it is for the Galaxy and for Cygnus A, with broken dust
clouds and, perhaps, multiple scattering. It is important also to
remember that the polarization images are smoothed by the seeing
weighted by the spatial distribution of $p\times\rm F_\lambda$, and so
in weakly polarized regions $\theta$ will be distorted towards its
value in nearby strong regions. 

   Note that the value of $\theta$ obtained with slit spectroscopy may
depend on the location of the slit and on the extraction window. In
general, the vectors are close to perpendicular to the radius, and so,
away from the nucleus, the measured $\theta$ will be close to
perpendicular to the slit orientation. This means that one must take
care in comparing a polarization to a radio axis in an NLRG like 3C
195, because the value of $\theta$ will depend on the details of how it
is measured.

\subsubsection{Spectropolarimetry} \label{813}

   For spectropolarimetry we oriented the slit to include the strongly
polarized regions shown in Figure 5d.  We extracted the spectrum and
polarization in the nucleus and in regions to the NE and SW, but in
this paper we only present spectra centered on the nucleus.
Subtraction of the galactic light is not enough to flatten $p_c$
(Figure 3). The corrected polarization rises to the blue and is about
$14\%$ at 3500 \AA. Also, broad H$\alpha$ and H$\beta$ are polarized
more strongly than the continuum; this might be due to a second
continuum component, FC2, as discussed in \S4.2; see also \S8.2.

\subsection{3C 33} \label{82}

   3C 33, a well-studied FR II radio source, is associated with a
galaxy having an EELR that extends far outside the galaxy. The velocity
field in the EELR has been mapped by \cite{bau90}, and an optical
synchrotron emission region is coincident with the southern hot spot
(\cite{mei97}).  The main H$\alpha$ axis is not close to the radio
axis; this is a case where the alignment effect is weak.  Our B-band
image is in Figure 5a.  A few emission lines, notably
[O~II]$\lambda3727$ and [Ne~III]$\lambda3869$, are in the B-band flux,
but the lines amount to no more than 25\% of the total and the image is
a good representation of the continuum. The lines themselves are
essentially unpolarized and do not contribute to the polarized flux.
The B-band contours are elongated in the central few arcseconds, with
an axis parallel to that of the H$\alpha$ image.  They become circular
outside a diameter of $8\arcsec$, so it is only close to the nucleus
that the continuum displays the elliptical shape, whereas the H$\alpha$
image is elliptical over its total extent, about $12\arcsec$.

   The polarized regions in Figure 5a are asymmetrical, but close to
the nucleus they lie roughly along the continuum and H$\alpha$ axis.
The NE polarization region is much larger than the one to the W. The
polarization shows a strong gradient, with a peak value 8.4\% in both
the NE and the W.  The polarization is low near the nucleus, where $p =
0.8\%\pm0.1\%$ within the central arcsecond. This value is low enough
that it could be affected by ISP, but the high galactic latitude
(-49.3$\deg$) suggests that the effect is small.  Three stars separated
from 3C 33 by 3.5\deg, 5.5\deg and 6\deg have polarizations 0.2\%,
0.4\%, and $<0.1\%$, respectively (\cite{mat70}). It is likely that the
ISP in the direction of 3C 33 is a few tenths of one percent, but we
ignore it because (a) in fact we do not discuss the polarization at the
nucleus in detail (see \S5.2), and (b) we cannot correct for it without
observations of stars close to 3C 33.  The weak ISP does not affect the
high polarization outside the nuclear region.

   The plot of $p\times \rm{F}_\lambda$ in Figure 5b is closely similar
to that of 3C~195 in Figure 5d. Both objects show separated peaks
differing in intensity by approximately 30\%, and a low-level extension
on one side.  The position angle $\theta$ rotates around the nucleus in
Figure 5a and stays roughly perpendicular to the radius. Thus 3C 33
provides another good example of a reflection nebula illuminated by a
point source. In the nucleus $\theta = 91\deg \pm 3\deg$. This is close
to parallel to the $p\times \rm{F}_\lambda$ axis rather than being
perpendicular to it, as in 3C 195. 3C 33 does not contain a clear
polarization axis, as 3C 195 does. The radio axis, however, is at
$19.5\deg$ which, like the axis in most of the NLRG, is within $20\deg$
of perpendicular to the nuclear $\theta$ (see \S10). Draper, Scarrott,
\& Tadhunter (1993) report imaging polarimetric observations of 3C 33
in V-band, which contains strong [O~III] lines and H$\beta$.  The
narrow lines are polarized less than the continuum, however, so the
Draper et al.  polarization image is dominated by the continuum and can
be compared with our B-band image. The results generally are similar,
and we confirm the centro-symmetric pattern they report.  However, we
find a stronger gradient, with low polarization at the nucleus.

   Spectropolarimetry of 3C 33 was made with the slit along the major
axis of the H$\alpha$ image of Baum et al. (1988; PA $62\deg$) as
shown in Figure 5b.  In Figures 1-3 we show spectra extracted from the
region of the NE peak in $p\times \rm{F}_\lambda$, 2$\arcsec$ from the
nucleus. This region has $\theta = 156\deg$, nearly perpendicular to
the slit direction, as expected for a resolved reflection nebula.

   3C 33 is an NLRG, but a weak broad line can be seen under H$\alpha$
in Figure 1a, and broad H$\alpha$ is the dominant feature in polarized
light, Figure 1b.  The starlight fraction is high (Table 3) and the
corrected flux in Figure 3 shows prominent broad Balmer lines.  The
polarization corrected for starlight is shown in Figure 3b.  In broad
H$\alpha$ $p_c$ rises to $\sim 25\%$; this is the highest value of
$p_c$ that we have measured.  This perhaps is not surprising, as it is
for an off-nuclear location where geometric dilution is expected to be
smaller than near the nucleus.  The continuum polarization is lower,
near 12.5\%, suggesting that there is a strong FC2 component. The
position angle shows no changes at H$\alpha$ so we can assume that the
nuclear continuum and H$\alpha$ are intrinsically polarized the same.
In this case we can follow the analysis of Tran (1995; see eqs
2.5-2.7), and we calculate $\rm{FC2/FC1} \sim 3$, and the intrinsic
polarization of FC1 and broad H$\alpha$, $p_{\rm s}$, is about 50\%.
That value is basically calculated at one wavelength, H$\alpha$, where
$p_c$ is a maximum. To obtain a feel for the validity of our analysis,
we plot in Figure 9 the spectrum of FC2 (the ``unpolarized flux
spectrum", see Tran 1995, Figure 1) with $p_{\rm s}$ held constant in
wavelength.  The top spectrum shows the corrected flux; this is the
same as the plot in Fig.  3a. The lower curves show spectra for $p_{\rm
s} = 0.3, ~0.5, ~\rm{and} ~0.7$. The fit can best be examined in the
blue wing of H$\alpha$, at 6500 \AA.  It is clear that $p_{\rm s} =
0.3$ undercorrects for FC2, and $p_{\rm s} = 0.7$ overcorrects. A value
between 0.4 and 0.6 appears to provide a good average correction over
the H$\alpha$ line.  The noise is higher in H$\beta$, but $p_{\rm s} =
0.5 \pm 0.1$ appears to be valid there also.

   Our value for the intrinsic polarization due to scattering, $p_{\rm s} =
50\% \pm 10\%$, is higher than values reported earlier. This is at
least partly due to the off-nuclear position of the measurement, but 3C
234 appears to have no geometric dilution (Fig 5e), and for it we found
$p_{\rm s} = 25\% \pm 3\%$ (T95).  In a group of 10 Seyfert 2 galaxies Tran
(1995) found a maximum value $35\% \pm 9\%$.  High values of $p_{\rm s}$,
like the one we find in 3C 33, in fact are expected in simple scenarios
of scattering and the low values more commonly found have required some
explanation (e.g.  \cite{mil90}).

\subsection{3C 105} \label{83}

   3C 105 is the only object in our sample that shows no evidence for
polarization, with $p < 0.8\% (3\sigma)$.  The galaxy is
exceptionally red, with strong stellar absorption lines (\cite{tad93}).
The Keck spectra are shown in Figure 10; the bottom panel is the
spectrum with a galactic template (corrected for reddening) subtracted.
There is no reliable indication of broad lines. 3C 105 has a typical FR
II radio morphology, with a weak core and prominent hot spots, and is
similar to 3C 357.

\subsection{3C 135} \label{84}

  3C 135 has several nearby companions (Figure 6), including one
$4\arcsec$ SW and a highly-distorted one $14\arcsec$ SW. The HST image
(\cite{dek96}) also shows a fine-scale extension from the nucleus to
the SW.  The polarization of 3C 135 is low; even after the strong
correction for host galaxian starlight it is only about 3\%, with a
large error. Eracleous \& Halpern (1994) did not find a broad component
of H$\alpha$ in 3C 135, and H96 similarly found narrow but not broad
P$\alpha$. However, although 3C 135 is weakly polarized, broad
H$\alpha$ and (perhaps) H$\beta$ are visible in $p$ and $p \times
\rm{F}_\lambda$.  3C 321 (Table 4) is similar in not showing broad
P$\alpha$ while showing broad H$\alpha$ in polarized light.

   In the polarization image of 3C 135 (Figure 6) the vectors are
stronger than $3\sigma$ in 10 bins near the nucleus.  The value
measured near the center is $p = 1.5\% \pm 0.2\%$ with $\theta =
159\deg \pm 3\deg$.  The image is taken in B band that contains
[O~II]$\lambda3727$ plus starlight in addition to the weak continuum.
The spectropolarimetry is noisy in B band but at 5500 \AA~the value is
$p=0.7\% \pm 0.1\%$ with $\theta = 142.1\deg \pm 4.8\deg$ (Table 3).
The values are consistent with a polarization that rises to the blue,
as is common in NLRG, and they are marginally consistent with $\theta$
being independent of wavelength. 

\subsection{3C 234} \label{85}

    The spectropolarimetry for 3C 234 has been discussed by T95. Figure 5
shows contour diagrams of F$_\lambda$ and $p \times \rm{F}_\lambda$.
$p$ increases to the East and has a maximum about $1.5\arcsec$ east of
the flux maximum. This gradient is compensated by the fall of
F$_\lambda$, so that $p \times \rm{F}_\lambda$ and F$_\lambda$ have
maxima in the same pixel. Figure 5f shows that the east-west asymmetries
in F$_\lambda$ and $p$ combine to produce $p \times$F$_\lambda$ that is
close to symmetric, although it is noticeably thinner north-south than
is F$_\lambda$.

   The spatial distributions of $p$ and $p \times \rm{F}_\lambda$ in 3C
234 are different from those in both the BLRG and the other NLRG.  The
polarization pattern is extended but does not show fans, and at the
nucleus $p$ is neither a minimum nor a maximum.  The pattern is
probably due to scattering of the nuclear continuum, but the geometry
seems to be different from that of the other objects.  The lack of
polarization fans suggests that we are looking inside the cone, but
we must be close to the edge because the broad lines are enhanced in
polarized light (much more so for 3C 234 than FSC2217+259).  It is hard
to explain the symmetry in $p \times$F$_\lambda$ in this case,
especially the lack of circumferential vectors to the north and south.

   The source $3.5\arcsec$ NE (Figure 5e) was recognized by Cimatti \&
di Serego Alighieri (1995) who stated that it is highly polarized at V
band. In our B-band polarization image no bin in that region shows $p >
2.5\sigma$. We integrated over a $1.1\arcsec$ square on the NE peak and
found $p < 1.4\%~(3\sigma)$. Hurt et al. (1999), using HST ultraviolet
imaging polarimetry, also fail to detect polarized light from this
region, and suggest that it is dominated by an old stellar population,
based on the red color.

\subsection{3C 321} \label{86}

   The B-band polarization image of 3C 321 is shown in Figure 5g. The
gradient in $p$ is stronger than it is for the other objects; and,
indeed, in the central region no individual bin has a measurable
polarization stronger than $2.5\sigma$.  The polarization is marginally
detected in a 4x4 bin $(0.8\arcsec$) average in the center: $p = 0.8\%\pm
0.3\%$.  $p$ rises to $9.3\%$ in the NW and to $7.2\%$ in the SE.

   The vectors are roughly circumferential, indicating scattering of
light from a point source. This pattern has already been noted by
Draper, Scarrott, \& Tadhunter (1993).  In an attempt to find the
source of the nuclear light, we analyzed the angle $\psi$ between the
local value of $\theta$ and the radius to an assumed center.  This
procedure has been used before by \cite{war97}.  When applied to 3C
321, it shows that $<|90\deg - \psi|>$ is  a minimum at the SE flux
peak, where it has the value $6.9\deg$, whereas at the NW peak it is
$14.3\deg$.  In other words, the vectors point (perpendicularly) more
to the SE component than to the NW component, and the source of nuclear
light appears to be in the southern component. This is consistent with
Filippenko's (1987) result that the compact radio source is in the SE
component.  3C 321 appears to be different from Cygnus A, where the
center of activity is hidden by dust and lies between the main nuclear
components.  Close inspection of the polarization vectors shows that
deviations are not random, but rather that the vectors are
systematically tilted away from the SE nucleus. This can be caused by
seeing, combined with the spatial gradients.

   Young et al. (1996) found broad H$\alpha$ in the polarized spectrum
of 3C 321 and, by using the Balmer decrement to estimate the reddening,
showed that the intrinsic luminosity was sufficient for the nucleus to
harbor a hidden quasar. The spectra we show in Figs 1-3 were made with
the slit along the optical symmetry axis and with an extraction window
$17\arcsec$ long, that includes essentially all the galaxy seen in
Figure 5.  Broad H$\alpha$ is weakly visible in the total flux in
Figure 1, but is prominent in the polarized flux and in the
galaxy-corrected flux (Figure 3).  Broad H$\beta$ and H$\gamma$ also
are visible. Figure 5 shows that the polarization varies widely over
the galaxy, so the measured values of $p$ and $\theta$ will generally
depend on the slit location and the extraction window.  The rapid rise
of $p$ to the blue in Figure 2 is partly but not completely due to
dilution by an old stellar population, as the increase of $p_{\rm c}$
in H$\alpha$ shows that FC2 is present also.  \cite{tad96} attribute
this extra continuum component to A stars. The spectrum of polarized
flux is steep in 3C 321, with $\alpha=-2.0$ (Figure 1a). This is the
steepest of the NLRG, and is as steep as 3C 382, which also has
$\alpha=-2.0$. In the UV the polarization rises to 40\% (\cite{hur99}).

   H96 did not detect broad P$\alpha$ in 3C 321, but in our
observations the broad lines are obvious in $p\times{\rm F_\lambda}$
(Fig. 1). This shows the importance of complementary observations. We
discuss this further in \S10.2.

\subsection{3C 357} \label{87}

   The HST image of 3C 357 (\cite{dek96}) shows dust lanes and a patch
of emission about $2\arcsec$ W of the galaxy; we call this region W.
On our Keck B-Band image (Figure 6) W is seen as the bulge in the
second-brightest contour (20\%). It lies close to the radio axis.  The
nucleus does not show any polarization, but in W four vectors survive
the $3\sigma$ cutoff, giving $p\approx 4\%$. The vectors are closely
perpendicular to the radius, suggesting a scattering origin for the
polarized light.  The spectroscopic slit at ${\rm PA=95\deg}$ was
centered on the polarized region, which is near the southern edge of
region W.  Both the nucleus and W show a great deal of starlight (Table
3) and in Figure 3 we show the galaxy-subtracted flux and polarization
of W.  (The flux is shown on a linear scale because it is so noisy.)
The narrow lines of W are red-shifted with respect to the nucleus, by
607 km/sec.

   Imaging polarimetry of W gives 4\% in B Band and spectropolarimetry
gives $p =1.9\% \pm 0.3\%$ at 5500 \AA.  The rise to the blue is due to
dilution by starlight, although narrow lines in the B band image should
reduce the B band polarization.  The galaxy-corrected polarization
$p_{\rm{c}} \sim 14\%$ in the continuum.  The noise is low in the
strong narrow lines [O~III]$\lambda\lambda4959,5007$ and (H$\alpha$ +
[N~II]) and $p_{\rm{c}}$ drops to near zero in the lines, indicating that
they are polarized much less than the continuum, and arise {\it in
situ}.

   From the nuclear spectra (not shown), an upper limit for $p(5500)$
in the nucleus is 0.4\% ($3\sigma$).  3C 357 is an unpolarized NLRG
like 3C~105 (\S8.3), but contains a fortuitously located scattering
region in the form of region W. It may be similar to the off-nuclear
scattering region in NGC 1068 (\cite{mil91}). There is no evidence of a
BLR in 3C 357, although our S/N is poor.  It would be worthwhile to
repeat these observations, to establish more clearly the presence or
absence of broad Balmer lines.  This is of importance in understanding
the unpolarized NLRG; is the observed lack of broad lines intrinsic, or
merely due to a lack of scattering material? The S/N of the current
data is too low for a definitive statement, but the apparent lack of a
scattered BLR is an intriguing suggestion that there may not be a BLR
in 3C 357.

\subsection{Cygnus A} \label{88}

   Cygnus A, the prototypical NLRG, is discussed by O97. The fact that
it has a reflection nebula and shows broad lines in scattered light
gives strong support to the unification by aspect of BLRG and NLRG.

\section{PA Rotations in the Broad Emission Lines}   

   Figure 2b shows intriguing rotations in $\theta$ across broad
H$\alpha$ in 3C 227 and 3C 445; there are smaller rotations in 3C 321
and Cygnus A, and similar effects in other lines in 3C 227.  Similar
rotations have been seen in Seyfert 1 galaxies
(\cite{goo94,martel96,smi97}) and in broad-absorption-line QSO
(\cite{ogl99}).  Evidently, such rotations are common in broad-line
objects. In this Section we describe the rotations in 3C 227 and 3C
445.

\subsection{3C 227} \label{91}

   Figure 11 gives an expanded view of the polarization of 3C 227,
showing (a,b) the Stokes parameters $Q_{\rm tot} = q\times \rm
F_\lambda$ and $U_{\rm tot} = u\times \rm F_\lambda$, and (c) the
position angle $\theta$.  To study the PA rotations in more detail we
decompose the polarized radiation into continuum and line components,
with $Q_{\rm tot} = Q_{\rm cont} + Q_{\rm line}$ and $U_{\rm tot} =
U_{\rm cont} + U_{\rm line}$.  $Q_{\rm cont}$ and $U_{\rm cont}$ are
approximated by the smooth curves in Figures 11a and 11b; they are
second-order polynomial fits to $Q_{\rm tot}$ and $U_{\rm tot}$ in
regions free of emission lines. The resulting $\theta_{\rm cont}$ is
the smooth curve in Figure 11c, which is shown together with
$\theta_{\rm tot} = 0.5\arctan(U_{\rm tot}/Q_{\rm tot})$.  In Figure
11d the smooth curve is again $\theta_{\rm cont}$ and the other is
$\theta_{\rm line} = 0.5\arctan(U_{\rm line}/Q_{\rm line})$.  We
emphasize that this construction depends critically on the exact curves
used for $Q_{\rm cont}$ and $U_{\rm cont}$, and a reliable
decomposition can only be made with high--S/N data.

   In the spectrum of 3C 227 (Fig 1) a weak broad feature is visible
at 5700-5900 \AA. This is a blend of He~I$\lambda5876$,
Na~I$\lambda\lambda 5890,5896$, Fe~VII$\lambda6087$ and, probably, other
lines. (See Thompson 1991 for a discussion of the He and Na lines.)
This feature is polarized and is seen in the curves of $Q$ and $\theta$
in Fig. 11. Similarly, the blue wing of H$\gamma$ contains a
contribution from broad He~II$\lambda4686$.  We shall refer to these
blends simply as He~I$\lambda5876$ and H$\gamma$.

   In Figures 11c and 11d it is clear that broad H$\alpha$,
H$\beta$, He~I$\lambda5876$, and possibly H$\gamma$, have a PA that
is different from that of the continuum.  H$\alpha$ and H$\beta$ are
rotated about $20\deg$ from the continuum, and $\theta$ is constant in
these lines to within about $10\deg$.  H$\gamma$ and He~I are rotated
from the continuum in the same direction as H$\alpha$ and H$\beta$ and
possibly by the same amount, although they are weak and the errors, due
mainly to uncertainties in the continuum levels, are large.

  The narrow lines are polarized less than the continuum and broad
lines, as is usual in these RG. [O~III]$\lambda5007$ appears as a
prominent spike in $\theta$, and formally is at $\theta \approx 5\deg$,
but we are suspicious of this value, because it occurs in only one
pixel. It might be due to noise, or to problems with the spectrum
extractions.  [S~II] appears to show up in $\theta$ and $\theta_{\rm
line}$, but this may be due to noise as the [S~II] signal is weak in $Q$
and $U$.

\subsection{3C 445} \label{92}

   Figure 12 shows the H$\alpha$ region of 3C 445 in the same construction
as for 3C 227, except that we used a linear fit to the continuum for
$Q_{\rm tot}$.  Note that in (c) and (d) the scales differ by a factor
of two.  The rotation in H$\alpha$ shows a striking red/blue
difference in 3C 445, one lacking in 3C 227. The crossover occurs at
about 6500 \AA, which is offset from the narrow-line systemic velocity
by 3000 km/sec. The red side in $\theta_{\rm line}$ (Figure 12d) is
roughly constant at $130\deg$ and is rotated by about $10\deg$ from
$\theta_{\rm cont}$.  This component clearly persists to 6820 \AA\ or
to the end of the red wing in total flux. Its total velocity width is
15,000 km/sec.  The blue side may persist to 6180 \AA, although the
noise near 6300 \AA ~makes this questionable. Its width is at least
7000 km/sec and could be up to 15,000 km/sec.  The value of
$\theta_{\rm line}$ clearly changes across the blue component, from
about $190\deg$ to about $170\deg$. The peak difference between the red
and blue components is about $70\deg$.

   The two values of $\theta_{\rm line}$ for H$\alpha$ suggest that we
are dealing with two components of the radiation, that are separated
in velocity and have different PA's. The blue component (B) is dominant
below 6500 \AA, and the red component (R) is dominant above 6500 \AA.
Inspection of the $p\times \rm{F}_\lambda$ curve (Figure 1b) shows that
H$\alpha$ has a pronounced break near 6500 \AA, and so such a
decomposition seems natural. B and R are separated in velocity by
$\sim 5000$ km/sec and are polarized $\sim 1\%$ with $\theta$'s that
are $\sim 70\deg$ apart.

\subsection{Interpretation} \label{93}

   The Seyfert 1 galaxies studied by Goodrich \& Miller (1994), Martel
(1996) and Smith et al. (1997) show a variety of rotations in the broad
lines, and several of their objects show rotations that are as large
and as complicated as those in 3C 445. These authors discuss this
phenomenon in terms of two or more components that we see via different
scattering regions.  The regions must lie fairly close to the BLR,
for the PA's to differ substantially.  Martel decomposes F$_\lambda$
into gaussians, and the best case is NGC 4151 which yields 7 broad
components.  He then identifies some of the components with structures
in the galaxy (a bar, radio and optical axes) on the basis of PA
coincidence.  We do not attempt a similar deconstruction for 3C 227 and
3C 445 here, but restrict ourselves to one component of H$\alpha$ for
3C 227, and two for 3C 445.

  In 3C 227 and 3C 445 none of the PA's is close to being either
parallel or perpendicular to the radio axis.  The polarization does not
result from simple scattering on a distribution that is symmetric
around the radio axis. What then is the source of the polarization?  In
3C 227 we evidently have different lines of sight into the nuclear
continuum source and the BLR. These light sources are spatially
separated by a few tenths of a parsec, so we can imagine that different
scatterers are responsible for the two different polarizations.
The LOS is not close to the axis of the dusty torus because the radio
source has FR II morphology with a weak core, and the inclination is
probably many tens of degrees. Thus, the required asymmetry is plausible.
A fraction of the H$\alpha$ light could be scattered and polarized by
dust clouds in or near the BLR, and it is possible that absorption
in aligned dust particles in front of (or mixed with) the H$\alpha$
clouds is responsible for some of the polarization.  The dust alignment
need not have any simple relation to the polarization of the continuum,
which is set much deeper inside the nucleus.

  3C 445 is more complex, in that we must account for two components of
H$\alpha$, but that also offers more constraints and may limit the
range of possible models. There is not enough information to invert the
polarimetry data and solve for the location and motion of the
H$\alpha$ clouds. However, we can investigate some simple limiting
cases and try to find plausible situations. We assume that the
system contains a BLR inside a dusty opaque torus, and that the
continuum source is at the center. The LOS skims the edge of the torus
and we have a limited view of the interior.

  We first note that the motions of the H$\alpha$ clouds cannot be
chaotic, since then the red-- and blue--shifted sources would be
spatially mixed, and the red and blue components would have the same
PA, contrary to observation. Hence the clouds must be in streaming
motion. This assumes that the velocity shifts are due to motions of the
emitting clouds, and not of the scatterers.  Further, the lines
cannot be broadened purely by thermal motions in clouds, for in that
case the electrons would produce equal red and blue shifts at every
velocity.  

   BLR clouds orbiting inside the torus, near the equatorial plane,
provide a scenario which can explain much of the observed asymmetry in
the polarization of H$\alpha$.  Consider the case of clouds orbiting in
circles in the equatorial plane.  The motions are not aimed at us, so
the observed high velocities must be due to scattering by clouds that
see the full red and blueshifts. In other words, the polarized
H$\alpha$ light that we see has been scattered by material that is
close to the orbital plane. This material can form the inner wall of
the torus.  If the rotating BLR clouds are optically thick they mainly
radiate back towards the ionizing source, but still, for circular
orbits, half of each emitting hemisphere is visible at the rim and the
full range of velocities can be seen in the scattered light.  An
appropriate combination of inclination angle and shape of the inner
wall of the torus can now provide red and blue components that have
different PA's.  See an example in Figure 13. In this example all
clouds move on circles in the same direction. In discussing Mrk 231,
Goodrich \& Miller (1994) describe a scenario that is similar to the
one we present here.  Our scenario only works over a restricted range
of inclination angles.  If the inclination is low the nuclear region is
entirely visible and we see a quasar. At high inclination we see an
NLRG with polarization perpendicular to the axis. If the view is near
the boundary of the torus the equatorial plane is partially visible
(Fig 13) and we see a BLRG with different PA's for the red and blue
wings of H$\alpha$.

   This simple model cannot be taken literally, for the situation
undoubtedly is more complex.  For example, \cite{nis98} have described
a scenario, based on X-ray spectroscopy, in which the BLR is in a
warped disk that is not coaxial with the accretion disk.  However, we
believe that our concepts still have some validity.  The clouds are in
streaming, not chaotic motion.  Optically thick clouds in equatorial
orbits provide a geometry that meets the observations, although the
3000 km sec$^{-1}$ offset to the blue is not readily accounted for.
Reverberation studies in Seyfert 1 galaxies (e.g.  \cite{kor91,kor95})
show that circular orbits are preferred over radial orbits, and this
provides an extra measure of support for our scenario.

\section{Discussion}                      

\subsection{Direct and Scattered Light} \label{101}

   The model in \S6 describes the BLRG radiation as a combination
of two components, a reddened direct ray and a polarized and reddened 
scattered ray. The relative amounts of the two rays fix the spectral index
and the amount and spectrum of the polarization. The NLRG have
more attenuation of the direct ray, so that the nuclear continuum and
broad H$\alpha$ are not directly seen. They show up in the scattered
light, and a weaker component, FC2, also becomes important. The
evidence for this is that $p_{\rm c}$ rises in H$\alpha$, indicating a
diluting unpolarized component in addition to the polarized continuum FC1.

   Observationally, the BLRG can be divided into 3 groups that have
rather different spectra and polarization. However, these should not be
thought of as separate classes of objects, since they are primarily
distinguished by differing extinction and inclination angle.

\noindent $(i)$   3C 382 is very blue, has low polarization, and broad
H$\alpha$ is prominent in total flux but not in polarized flux.  In its
brighter phases, this object is a quasar. The low polarization of
H$\alpha$ can be understood if we have a direct, unattenuated view of
the BLR, so that the scattered (polarized) light from the BLR is weak
in comparison to the direct ray. Another way of looking at this is to
imagine that all BLRG have a more-or-less standard spectrum of
$p\times{\rm F_\lambda}$, as exemplified by 3C 227. 3C 382 additionally
has a strong unpolarized component of H$\alpha$ that dilutes the
polarization in the line. This combination leaves the position
angle unchanged in the line, as required by Figure 2b. 

\noindent $(ii)$   3C 227 and 3C 445 are highly polarized in the blue.
The broad lines are equally prominent in total and polarized flux.  PA
rotations in H$\alpha$ show that there are multiple lines of sight to
the BLR.  We have a view deep into the illuminated cone, but may not
directly see the nuclear continuum source or the BLR.  The view
probably is close to the edge of the torus. These two objects are
analogous to the Type 1 Seyferts that have PA rotations, but the RG
also have powerful radio jets.

\noindent $(iii)$ 3C 109 and FSC 2217+259 are very red, highly
polarized, and the polarization of broad H$\alpha$ is the same as that
of the continuum. They are closely similar except for the spectral
index of the continuum (Figure 7).  FSC 2217+259 is a quasar seen
through about 3 magnitudes of dust extinction (according to \S7.4.1),
with a scattered component in which the continuum and broad lines are
polarized the same.  The net polarization $(p_c)$ is high and rises
strongly to the blue; this is due to the reduction in dilution by the
heavily-reddened direct component.  3C 109 has been interpreted as
being polarized by selective absorption in aligned dust grains, but in
view of its similarity to FSC 2217+259 we think it may also have a
scattered component. Observations into the UV would be useful for both
these objects.

   In any event, 3C 109 and FSC 2217+259 are quasars seen through
dust.  We are looking deep into the torus, and may have a direct
(but attenuated) view of the BLR and the nuclear continuum source.  FSC
2217+259 does not have a compact optical nucleus like the other BLRG,
and its image of $p \times \rm{F}_\lambda$ resembles that of 3C 234.

\subsection{Unification} \label{102}

  Our work strongly supports the hidden nucleus hypothesis for NLRG
and the unification by aspect of many FR II NLRG, BLRG and quasars,
in four ways.

\noindent$(i)$ Reflection nebulae, in the form of fans of polarized light,
are seen in some NLRG. The polarization vectors are circumferential
and accurately define a central hidden continuum source. The fans are
roughly co-spatial with the EELR, and the radio axis is usually inside
the fans.

\noindent$(ii)$ Broad H$\alpha$ is seen in 6 of the 7 polarized NLRG,
showing that they contain a hidden BLR.  Our high-quality spectra show
this feature directly, but it is much more prominent in polarized light.
The seventh polarized object, 3C 357, is the weakest and is 
unreliable because the atmospheric A band falls on the red wing of
H$\alpha$.

\noindent$(iii)$ The intrinsic luminosity of the BLRG in our sample,
and also of 4 NLRG, 3C 227 (Prieto et al. 1993), 3C 234 (T95), 3C 321
(\cite{you96}) and Cyg A (O97), has been investigated in detail.  It is
sufficient to put all of them in or close to the quasar class.

\noindent$(iv)$ In the BLRG the outer reflection nebulae must exist, as
they do for the NLRG, but they have not been seen.  However, in BLRG
the nuclear continuum and broad lines swamp the weaker light from the
reflection nebula. A measure of this difference can be gained from
Table 3, where the galaxy fractions $f_g$ show that the starlight (and
presumably the outer scattered light) is weaker in the BLRG than in the
NLRG by a factor of 3 to 10 or more. A further reason why the reflection
nebulae might be hard to see in the BLRG is foreshortening due to the
smaller inclinations. Although the highest-redshift object is 3C 109,
a BLRG, the other BLRG have redshifts in the range of the NLRG, and 
angular size is not a major factor in the visibility of the EELR.

    Another commonly-used aspect indicator is R, the ratio of the radio
flux density in the core to the lobe flux density. Log R at 6 cm
wavelength is listed in Table 5. The mean value of log R for the BLRG
is --1.1 and for the NLRG it is --2.1. The difference is highly
significant and shows that the BLRG radio axes are closer to the LOS
than the NLRG axes, on average.

   These results justify the common scenario of an occulting torus that
hides the nucleus from direct view but allows collimated nuclear light
to be scattered into our LOS. From other aspects the NLRG would look
like BLRG and quasars.  Our results suggest that most polarized NLRG
contain a hidden BLR.  However, we cannot generalize about the fraction
of unpolarized objects that might show broad lines because we have only
one unpolarized NLRG.  We also cannot generalize to find the
statistical occurrence of broad lines in all NLRG, for the 
sample is biased towards objects known in advance to be polarized.  

   Broad lines should be seen in the IR if the dusty torus picture is
correct. H96 looked for P$\alpha$ in a complete sample of 11 FR II
radio sources, including 1 quasar, 1 BLRG, and 9 NLRG.  Their sample
includes 3C 273 which usually is not called FR II because it is not a
classical radio double. However, we keep it because it has high
luminosity and one visible relativistic jet with an outer lobe.  One
NLRG, 3C 319, is dominated by starlight and has no apparent H$\alpha$;
it was not observed in the IR but we keep it in the sample, which
consists of all FR II 3CR objects in the ranges of flux, redshift,
latitude, right ascension and declination chosen by H96.  We have
changed their designation of 3C 234 from BLRG to NLRG.  Three of their
NLRG, 3C 135, 3C 234, and 3C 321, are in our sample. Table 4 summarizes
the combined data sets. Y and N simply indicate whether there is or is
not evidence for high polarization, broad H$\alpha$ and broad
P$\alpha$.  We have added the polarization value (N) for 3C 236 from
\cite{imp91}, because although a galaxy-subtracted value is not
available, the measured value is low, $0.2 \pm 0.3\%$. Table 4 contains
14 NLRG, of which 10 show evidence of broad lines and 4 do not. There
is a strong correlation between polarization and the existence of broad
lines: 6 of the 7 NLRG with high $p_c$ show broad lines, and the 2 with
no measurable continuum polarization do not show polarized broad
lines.  This reinforces the picture seen in our data alone, that
polarized objects will show broad lines, either in polarized light or
in the IR.

   H96 had a complete, if small sample, and the combination of our
spectropolarimetric results with their near-IR results is more powerful
than either technique alone.  This produces a result that is
correspondingly more robust. While H96 did not see broad P$\alpha$ in
3C 135 and 3C 321, we have shown that there is indeed a BLR in
these objects. This brings the number of objects known to have a BLR to
8 in the complete H96 sample of 11 FR II objects, or 6 of the 9 NLRG.

   \cite{tad97} have made a polarization survey of a complete sample of
radio galaxies, of which 6 are FR II NLRG. Three of the 6 are
polarized, but none shows evidence of broad lines. This seems to
contradict our result that polarized NLRG show broad lines, but limits
on the strength of the broad lines are not available for their
objects.  We believe that a sensitive survey of a larger complete
sample is warranted.

   The question of whether the unpolarized FR II NLRG contain a BLR is
important, and unanswered.  In particular, the lack of evidence for
broad lines in an object should not be construed as showing that it has
no BLR. There may be no signs of broad lines because there is no BLR, or
because there are no appropriately placed scattering regions.  3C 357
is essentially unpolarized except for a small region (3C 357W) where
the HST image shows an off-nuclear emission region.  The polarization
there is approximately perpendicular to the line to the center (Figure
6), which suggests that it sees a central continuum
source.  It does not appear to scatter any broad lines, but the low S/N
does not allow a firm conclusion to be drawn about the existence of a
BLR in the nuclear region.

  3C 105 is unpolarized but still is a powerful FR II radio galaxy with
a strong NLR, and it should have a bright central engine. In this case
the lack of continuum polarization may be due to a lack of scattering
material. The lack of observed broad lines could be due to the absence
of a BLR, but alternatively it could also be due to a lack of
scattering material.  The referee observes that the apparent lack of
scattered nuclear light could alternatively be due to obscuration of
the scatterers; this is possible but seems unlikely because the
obscuring material would have to cover a large area of the ISM.  A
polarization image of 3C 105 could be used to search for off-nuclear
scattering regions which may have fallen outside our slit.  R.A.
Fosbury points out (private communication) that it also is conceivable
that the central black hole is not currently being fed, so that the
nuclear continuum source and BLR are weak, but the long time scales
associated with the radio source and the NLR keep them going. As far as
we are aware, there are no VLBI observations of 3C 105, so there is no
structural information on scales smaller than a few hundred pc.

\subsection{The Radio Connection} \label{103}

   In powerful RG the radio axis is assumed to mark the dynamical axis
of the central black hole and accretion disk, and it usually is assumed
that any obscuring torus has the same axis. In FR II objects the radio
jets are nearly straight and the projected axis is well-defined.  In
Table 5 we give PA for the radio axis, $\theta$ for the optical
polarization, and $\Delta\theta$, the acute angle between them. The
value of $\theta$ for 3C 33 is for the nucleus whereas in Table 3 it is
for the NE lobe, corresponding to the spectrum in Figure 1; and for 3C
321 it is for the SE flux peak (Figure 5g) whereas in Table 3 it is for
the wide spectrum extraction.  In the past much attention has been paid
to the relation between $\theta$ and the radio PA, and in particular to
the relative orientation of the radio axis with the extra-nuclear UV
and emission-line light (alignment effect).  In 3C 195, 3C 321 and
Cygnus A the radio jet is well inside the polarization fans and the
torus axis can be taken as coincident with the radio axis, although the
torus is far from a smooth circular structure and the illuminated cones
must be irregular. In 3C 33 the polarized fans are not coaxial and the
collimation mechanism appears to give different illumination patterns
on the two sides, even though the radio structure is nearly straight
across the nucleus.  The inner 6\arcsec~ of the $p\times \rm{F}_\lambda$
image (Figure 5b) is nearly symmetric but the large weak region to the
N is not repeated in the S. In 3C 33 the $p\times \rm{F}_\lambda$ axis
is far ($50\deg$) from the radio axis.

   The important feature of the polarization image in NLRG is that it
shows the reflection nebula. At the nucleus itself the polarization
becomes weak but in all cases it is within $24\deg$ of being
perpendicular to the radio axis (Table 5).  We think this comes about
because, when there is sufficient symmetry at the apex, the
contributions add up to give an orientation perpendicular to the cone
axis, which is close to the radio axis. In
3C 33 the alignment effect is one-sided, but the nuclear polarization
is still roughly perpendicular to the radio axis. In this case the
polarization rotates strongly and is weak at the nucleus (Figure 5a) so
that observations must have high S/N and high angular resolution to be
reliable.

  \cite{ant84} noted that the BLRG are mostly polarized with $\theta$
roughly parallel to the radio axis. From Table 5 we see that the BLRG
differences do not cluster near $0\deg$ or anywhere else.  However,
there are none near $90\deg$, and the median value of $\Delta\theta$
for the BLRG is $31\deg$, whereas for the NLRG the mean is $77\deg$.
See also Hurt et al. (1999) for NLRG measured in the UV; they get
essentially the same result.  The geometry is simple for the NLRG,
because the bright central source is occulted and the scattering takes
place far from the nucleus.  With the BLRG we have a view deep into the
ionization cone, but perhaps not into the nucleus itself.  The
polarization is predominantly produced by scattering, and if the ISM is
irregular a range of $\theta$ is to be expected.  The one object that
looks like a quasar, 3C 382, has $\Delta\theta = 11\deg$, showing that
its polarization is close to parallel to the radio axis.  In
this case we apparently do have a direct view of the nuclear continuum
source, and the various models that have been proposed to explain the
polarization, typically involving an equatorial disk, may be
appropriate.

   The FR II radio galaxies have a powerful radio jet that is
indicative of magnetic field and energetic particles. Hence we might
expect that synchrotron radiation could play a role in the optical
emission and polarization. None of our objects shows direct evidence of
nuclear synchrotron radiation. However, we believe that they are
obscured or mis-directed quasars. Thus if (radio-loud) quasars
generally do contain a mis-directed blazar, then our objects should
also, and there should be a component of synchrotron emission in the
continuum radiation.  The direct synchrotron beam is narrow and only
seen close to the axis.  However, in misdirected objects the beam
should illuminate the ISM and there should be a narrow bright scattered
component. The only possible sign that we have of this is the polarized
off-nuclear emission region in 3C 357, and the fact that it is close to
the radio axis lends weight to this interpretation. However, the S/N is
low in this case, and better observations are needed before this
possibility can be discussed.

   In their discussion of Cygnus A, O97 noted that the larger and more
highly polarized fan is on the west side, and that this is the front
side of the galaxy, as determined by radio observations. They commented
that this is consistent with increased forward scattering on dust
particles.  We do not have the conclusive information provided by VLBI
for the other NLRG with fans, but still can compare the size and
strength of the fans with other indicators. We have 3 potential
indicators of sidedness: radio structure, polarization fans, and
velocity. Table 6 shows the results. Column 2 gives the side with the
larger and stronger polarization fan. As can be seen in Figure 5, these
are the regions which have extensions in $p\times{\rm F_\lambda}$,
although the peak values of $p\times{\rm F_\lambda}$ are on the
opposite side. We have individual spectra for the two sides and the
blue-shifted side is noted; this side is in front if the material is
out-flowing. We also attempted to establish the side in front with a
radio image.  In Cygnus A this is unambiguously determined by VLBI. In
the others we chose the side that has the more complete and stronger
jet, but in 3C 33 we were unable to choose.  It appears in Table 6 that
the correlations are no more than would be given by chance, and that
the Cygnus A result cannot be generalized.

\section{Summary and Conclusions}

1. We have sensitive Keck spectral and imaging polarimetry on 13 FR II
radio galaxies -- five BLRG and eight NLRG.  One NLRG is unpolarized
and the others have continuum polarizations ranging from 0.7\% to 9.6\%
at 5500 \AA~(rest frame).  When $p$ is corrected for galactic starlight
the polarizations rise to 15\%, and in 3C 33 a further correction for
FC2 yields a value $p_{\rm s} \sim 50\%$ for the intrinsic polarization of
the scattered nuclear light.  High values like this are expected from
scattering in an optically thin cone.

2.  Broad H$\alpha$ is visible in six of the seven polarized NLRG.  In
five of them the broad lines can be directly seen in the spectrum of
total flux, but they are more prominent in polarized flux.
In all the polarized NLRG except 3C 357, and perhaps 3C
135, the polarization of broad H$\alpha$ after correction for galactic
starlight is higher than in the neighboring continuum.  This is
evidence for an additional polarization-diluting continuum component,
called FC2. Its nature is unknown but it could be due to hot stars or
brehmsstrahlung.

3. The combination of our work with that of H96 shows that at least 6/9
of a complete, volume-limited sample of FR II NLRG have broad lines, seen
either in polarization or P$\alpha$.

4. $\theta$ (measured at the nucleus) is within 24\deg of perpendicular
to the radio axis for all the NLRG; the mean value of $|\Delta\theta| =
77\deg$.  The BLRG are not close to parallel. The median value of
$|\Delta\theta| = 31\deg$, and only one (3C 382) has $|\Delta\theta| <
25\deg$.

5.   The B band NLRG polarization images show a symmetric polarization
pattern characteristic of a bipolar reflection nebula.  These objects
contain a BLR and a continuum source which is hidden from us by dust
but revealed by the polarization spectra and images. From directions
within the scattering cone they would look like BLRG or quasars. Hence,
the obscuring torus paradigm unifying Seyfert 1 and 2 galaxies also
applies to FR II radio galaxies.

6. A large fraction of polarized NLRG contain a BLR, but we do not know
if the correlation between $p$ and the BLR is intrinsic or merely due
to the existence of appropriately located scattering material that
produces the polarization and also allows the BLR to be recognized.
Conversely, we have one unpolarized NLRG that does not show broad
lines.  We do not know if the BLR is missing or if there is just a
lack of scattering material.  We are unable to generalize
these results to find, for example, the fraction of NLRG that are
polarized, as our sample is strongly biased towards objects known in
advance to be polarized. 

7.  There is a hint in the low S/N data for 3C 357 that there is no
BLR, because the west emission region (W) appears to partly consist of
scattered nuclear continuum light, but there is no broad H$\alpha$.
However, this suggestion  is not correct if W is scattering a beamed
synchrotron component.  There probably is a variation in the luminosity
of the BLR in these RG, and that, together with the aspect and
extinction, can explain the gradation in broad H$\alpha$ seen in
Figures 1 and 3. We follow \cite{ant90} in defining an NLRG as one in
which the BLR (and continuum source) is mainly seen by scattering, and
a BLRG as one dominated by direct nuclear light.  By this definition,
3C 234 is an NLRG although broad lines are easily visible in total
flux.  We suggest that trying to draw careful distinctions here is
unimportant, and that it is the geometric and physical properties in
the nucleus that are of interest.

8. We propose a model for the BLRG in which the spectra are composed of
differing fractions of direct and scattered light, with reddening.  The
model explains the general blue shape of the polarization spectra, but
does not fit 3C 109 and FSC 2217+259 very well. These are the two red
BLRG with large extinction on the direct ray.

9.  3C 227 and 3C 445 have marked rotations in $\theta$ in the broad
lines.  We interpret this to mean that there are multiple lines of
sight to the BLR. We have a partial view deep into the nucleus but
probably near the edge of the torus.  We suggest that the broad-line
clouds are in streaming, not chaotic motion, and we describe
a simple scenario of clouds in equatorial orbits that can explain much
of the observed rotation phenomena in 3C 445.

10. 3C 382 was very bright and was a quasar at our observing epochs. At
other epochs it has been several magnitudes weaker and had an
easily-visible host galaxy. In this case the names quasar and BLRG have
little physical distinction other than to define whether or not the
nuclear source is brighter than the stars.
 
\acknowledgments

   We are grateful to J. Beverly Oke and Judy Cohen, first for building
the LRIS and second for the many discussions we had about its use.  We
are grateful to the instrument specialists at Keck, Tom Bida and Randy
Campbell, for their support that often went well beyond the call of
duty. We thank I. Browne and M. Marcha for telling us about the high
polarization of FSC 2217+259 in advance of publication, and R. Fosbury
and J. Vernet for suggestions concerning dust and other matters.  We
thank R. Antonucci, L.  Armus, J. Baker, L.  Ferrarese, D.  Hines, P.
McCarthy and S. di Serego Alighieri for useful discussions.  We are
grateful to R.  Fosbury and S.  di Serego Alighieri for assistance with
the observations, and thank the referee, Bev Wills, for many useful
comments and suggestions.  We used the vital NASA Extragalactic Data
Base (NED) in this research. The W.M.  Keck Observatory is a scientific
partnership between the University of California and Caltech, made
possible by the generous gift of the W.M.  Keck Foundation and support
of its president, Howard Keck.  This work was partly supported by NSF,
Grant AST-9121889, and research by HDT at LLNL was supported by DOE
under contract W7405-ENG-48.


\break

\figcaption{(a) (left panels): Total flux spectra of 11 FR II radio galaxies.
The top 5 objects are BLRG, the bottom 6 are NLRG. Spectra are arranged 
in order of decreasing visibility of broad H$\alpha$. Note the reddening 
sequence of the BLRG. The NLRG continua are dominated by starlight from the
host galaxies. The logarithmic ordinate has units 
$10^{-15}$ erg sec$^{-1}$ cm$^{-2}$ \AA$^{-1}$. (b) (right panels): Polarized 
flux $\equiv p\times \rm{F}_\lambda$. Units are $10^{-17}$ erg sec$^{-1}$ 
cm$^{-2}$ \AA$^{-1}$. Note that polarized broad lines appear in both BLRG and 
NLRG.  The gaps in 3C 445 and 3C 109 spectra are caused by the
dichroic reflector at the Hale Telescope. \label{fig1}}

\figcaption{(a) (left panels): Fractional polarization $p$ of the 
radio galaxies. In most cases, $p$ increases to the blue and falls in the 
narrow emission lines. Increased polarization across broad H$\alpha$ is an
indication of diluting continuum sources. (b) (right panels): Position angle 
$\theta$ of the polarization. $\theta$ is fairly constant with wavelength for 
most objects. However, note the rotations across broad H$\alpha$ in 3C 227 and
3C 445.\label{fig2}}

\figcaption{Galaxy-corrected spectra. See Table 3 for the galaxy fraction
at 5500\AA. (a) (left panels): Flux F$_{\lambda,c}$ F$_{\lambda,c}$ (plotted
logarithmically except for 3C 357W; see text). Note that broad 
H$\alpha$ is more easily visible after galaxy subtraction. (b) (right panels): 
galaxy-corrected polarization $p_{\rm c}$. Polarization rises to the blue even
after galaxy correction. The polarization increases at broad H$\alpha$, an
effect attributed to FC2 (see text). \label{fig3}}

\figcaption{BLRG polarization images, B band. Note that the polarization is
concentrated in an unresolved nucleus. All images have north up
and east to the left. Total flux contours for 3C 382, 3C 109 and 3C 227 are 
0.2, 0.4, 1, 5, 50\% of peak; and for FSC 2217+259 they are 5, 10, 20, 50\%
of peak.  These images are binned 2x2 from the original data and the
scale is $0.43\arcsec$ per bin, or $2\arcsec$ per tick. A 1$^{\prime\prime}$
polarization vector represents 5, 17, 10, and 20\% polarization for 3C 382, 
3C 109, 3C 227, and 2217+259, respectively . The vectors are suppressed for 
$P < 3\sigma$ ($3.5\sigma$ for 3C 382).  The central pixel of 3C 382 is 
saturated. The axis of the radio structure is marked R.\label{fig4}}

\figcaption{NLRG polarization images, B band. In contrast to the BLRG,
the polarized emission is extended and shows a fan structure in some cases.
Top: contour diagrams of F$_\lambda$ and the polarization vectors, with radio 
axis marked R. Bottom: contour diagrams of $p\times \rm{F}_\lambda$ with slit 
marked S. Scale as in Figure 4. Contours, in percent of the peak, 
are a(1.5,3,8,20,50), b(5,15,30,50,80), c(1,2,5,15,50), d(10,20,40,65,85), 
e(0.5,1,2,10,50), f(1,2,5,15,50), g(1,2,4,10,20,50), h(10,25,40,70,90). 
A 1$^{\prime\prime}$ polarization vector represents 17, 25, 33, and 20\% 
polarization for a, c, e, and g, respectively. \label{fig5}}

\figcaption{NLRG polarization images, B band. Scale as in
Figure 4. Total flux contours, in percent of the peak, are 3C 135(1,2,5,15,50),
3C~357(3,5,10,20,50). A 1$^{\prime\prime}$ polarization vector represents 
10\% and 8\% polarization for 3C 135 and 3C 357, respectively. \label{fig6}}

\figcaption{(a) Broad-line Balmer decrement vs galaxy-corrected continuum
flux ratio for the single-screen model in \S6.1.  Four reddening lines
for recombination Case B are shown (solid), with ${\rm\alpha_o = -0.5,
0.0, 0.5, and 1.0}$.  Lines of constant extinction are shown dotted;
the extinctions run from ${\rm A_V} = 0$ to 3.5 mag.  (b)
Galaxy-corrected polarization at 5500 \AA~(rest) vs flux ratio for the
two-component model in \S6.2.  Solid and dotted lines represent the
same quantities as in (a).  (c) Balmer decrement vs flux ratio for the
two-component model.  Solid and dotted lines represent the same
quantities as in (a).\label{fig7}}

\figcaption{3C 195 polarization position angle $\theta$. Note the line
of constant position angle at $\theta\simeq 41\deg$ (yellow), which
marks the symmetry axis of the reflection nebula. The $\theta$ rotation
is apparent in the north, but less so in the south.  The data are
suppressed for $p<3\sigma$.  Colors are coded as follows:
indigo$=84\deg$, blue$=92\deg$, green$=111\deg$, light green$=119\deg$,
yellow$=137\deg$, and deep red$=155\deg$. \label{fig8}}

\figcaption{Top spectrum: 3C 33 corrected for Galactic reddening
and for a population of old stars. Bottom spectra show the results of
correcting for FC2 with $p_{\rm s} =$ 0.3, 0.5 and 0.7 (see text). 
The bottom 3 spectra are binned to 20 \AA.\label{fig9}}

\figcaption{Top: original, uncorrected spectrum of 3C 105. Bottom: dereddened,
 galaxy-subtracted spectrum of 3C 105.\label{fig10}}

\figcaption{3C 227 separated into continuum and emission line
components.  (a) $Q_{\rm tot}$ and $Q_{\rm cont}$. (b) $U_{\rm tot}$
and $U_{\rm cont}$.  (c) $\theta = 0.5 \arctan(U_{\rm tot}/Q_{\rm
tot})$ and $\theta_{\rm cont} = 0.5 \arctan(U_{\rm cont}/Q_{\rm
cont})$. Note the strong rotations across the broad lines. (d) 
$\theta_{\rm line}$ and $\theta_{\rm cont}$ (see text).
Tick marks indicate H$\gamma$, H$\beta$, [O~III]$\lambda\lambda4959,5007$, 
HeI$\lambda5876$, H$\alpha$, and [S II]$\lambda\lambda6716,6731$.\label{fig11}}

\figcaption{3C 445 separated into continuum and emission line 
components, as in Figure 11. Tick marks indicate H$\alpha$ and
[S II]$\lambda\lambda6716,6731$.\label{fig12}}

\figcaption{Schematic diagram showing the partially-visible equatorial
plane of 3C 445.  Broad H$\alpha$ clouds orbit on a circle inside an
elliptical torus. Red and blue shifted components (R and B) are
scattered and seen with a PA difference of about 30\deg.  When
integrated over the visible sector of the torus, red and blue
components have the same intensity and fractional polarization, with
different position angles.\label{fig13}}


\begin{deluxetable}{llccccl}
 \tablecaption{Low Redshift Radio Galaxies}
 \tablehead{\colhead{IAU} & \colhead{Name} & \colhead{Type\tablenotemark{a}} &
            \colhead{z\tablenotemark{b}} & \colhead{${\rm b^{II}(deg)}$} & 
            \colhead{ISP$_{\rm max}$(\%)} & \colhead{Refs}}      
\startdata
0106$+$130 & 3C 33    &  N  & 0.0592 & $-$49.3 & $<$0.3 & 1, 2, 18  \nl      
0404$+$035 & 3C 105   &  N  & 0.0890 & $-$33.6 &  1.3 &      \nl
0410$+$110 & 3C 109   &  B  & 0.3056 & $-$27.8 &  2.3 &  3, 4, 5, 6 \nl  
0511$+$008 & 3C 135   &  N  & 0.1274 & $-$21.0 &  1.1 &      \nl
0806$-$103 & 3C 195   &  N  & 0.1100 &    12.0 &  1.1 &  7, 8 \nl     
0945$+$076 & 3C 227   &  B  & 0.0862 &    42.3 & $<$0.3 &  1, 3, 8, 9 \nl 
0958$+$290 & 3C 234   &  N  & 0.1848 &    52.7 & $<$0.3 &  1, 3, 4, 6, 7, 10, 18 \nl
1529$+$242 & 3C 321   &  N  & 0.0961 &    53.9 &  0.4 &  2, 4, 7, 11, 18 \nl   
1726$+$318 & 3C 357   &  N  & 0.1661 &    30.6 &  0.5 &      \nl
1833$+$326 & 3C 382   &  B  & 0.0579 &    17.4 &  0.7 &  1, 3, 9 \nl     
1957$+$406 & Cyg A    &  N  & 0.0561 &     5.8 &\nodata &  12, 13, 14, 15, 16, 18\nl 
2217$+$259 &FSC 2217+259&  B  & 0.0850 & $-$25.4 &  0.8 &  17     \nl    
2221$-$023 & 3C 445   &  B  & 0.0562 & $-$46.7 &  0.4 &  1, 3, 6, 9 \nl  
\enddata

\tablenotetext{a}{N$=$NLRG, B$=$BLRG}
\tablenotetext{b}{Redshifts from NED}
\tablecomments{References.--(1) Antonucci 1984; (2) Draper et al 1993; 
(3) Rudy et al 1983; (4) Cimatti et al 1993; (5) Goodrich \& Cohen 1992; 
(6) Brindle et al 1990; (7) Cimatti \& di Serego Alighieri 1995; 
(8) di Serego Alighieri et al 1997; (9) Corbett et al 1998; 
(10) Tran et al 1995; (11) Young et al 1996; (12) Goodrich \& Miller 1989; 
(13) Tadhunter et al 1990; (14) Jackson \& Tadhunter 1993; 
(15) Antonucci et al 1994; (16) Ogle et al 1997; (17) Marcha et al 1996;
(18) Hurt et al 1999.}
\end{deluxetable}


\begin{deluxetable}{lccccc}
 \tablecaption{Observations}      
 
\tablehead{Object &\multicolumn{3}{c}{SPECTROSCOPY} & \multicolumn{2}{c}{IMAGING} \nl  & \colhead{Date} & \colhead{Exp(sec)} & \colhead{Slit PA}& \colhead{Date} & \colhead{Exp(sec)} \nl }
\tablecolumns{6}
\startdata
3C 33  &   10 96   & 1500   &  62    &   10 96  &1200 \nl
3C 105 &   12 95   & 2880   &  29\tablenotemark{a}    & \nodata      & \nodata \nl
3C 109 &11 90\tablenotemark{p}~, 1 91\tablenotemark{p}& 8400  &  90\tablenotemark{b}    &   12 95  & 2400 \nl
3C 135 &10 94, 12 95&3600, 5280 & 165, 0  &   10 96  & 2400 \nl
3C 195 &   12 95   & 6720   &  35\tablenotemark{a}    &12 95, 4 96& 2400, 2400 \nl
3C 227 &   12 94   & 1920   & 120    &   12 95  & 1440 \nl
3C 234 &   10 94   & 4800   &  90    &12 95\tablenotemark{m}~, 4 96 & 2880, 2400 \nl
3C 321 &    5 96   & 3600   & 130    &    4 96  & 1200 \nl
3C 357 &    5 96   & 3600   &  95    &    4 96  & 2400 \nl
3C 382 &    5 96   & 1920   &  90    &    4 96  & 120\tablenotemark{s} \nl
Cyg A  & 8 94, 10 96&3600, 7920&  35, 101&   10 96  & 3600 \nl
FSC 2217+259 &    7 95   & 1200   & 140    &   12 95  & 240 \nl
3C 445 &    7 94\tablenotemark{p}   & 3600   & 159\tablenotemark{b}    & \nodata & \nodata \nl
\enddata
\tablenotetext{a}{slit width $1.5^{\prime\prime}$} 
\tablenotetext{b}{slit width $2^{\prime\prime}$} 
\tablenotetext{m}{possibly contaminated by moonlight} 
\tablenotetext{p}{Palomar} 
\tablenotetext{s}{saturated}
\end{deluxetable}


\begin{deluxetable}{lcccccc}      
 \tablecaption{Polarization}
 \tablehead{\colhead{Object} & \colhead{Bin\tablenotemark{a}} & 
            \colhead{Win ($^{\prime\prime}$)\tablenotemark{b}} &
            \colhead{$p$\tablenotemark{c}$~(\%)\pm$} & 
            \colhead{$\theta$\tablenotemark{d} $\pm$} &
            \colhead{$f_g$\tablenotemark{e}} & 
            \colhead{$p_{\rm c}$\tablenotemark{f} (\%)}}   
\startdata
3C 33NE & 8 & 2.4 & 3.1 0.4 & 150.5 2.2 & 0.74 & 11.9\nl
3C 105  & \nodata & 8.6 & 0.0 0.3 & \nodata   & 0.78 &$<$2.7\nl
3C 109  & 4 & 5   & 6.8 0.2 & 170.5 0.6 &\nodata&\nodata \nl
3C 135  & 8 & 2.2 & 0.7 0.1 & 142.1 4.8 & 0.85 &  4.7\nl
3C 195  & 4 & 1.9 & 2.3 0.1 & 128.1 0.9 & 0.62 &  6.1\nl
3C 227  & 4 & 3.7 & 2.3 0.1 &  41.6 0.7 & 0.30 &  3.3\nl
3C 234  & 4 & 2.2 & 9.6 0.2 & 156.3 0.4 & 0.35 & 14.8\nl
3C 321  & 6 & 17.2 & 1.1 0.1 & 40.4 2.7 & 0.66 &  3.2 \nl
3C 357w & 8 & 2.2 & 1.9 0.3 &  24.0 6.6 & 0.87 & 14.6\nl
3C 382  & 4 & 5.4 & 1.0 0.0 &  63.0 0.4 &$<$0.10& 1.0\nl
Cyg A   & 4 & 7.7 & 2.3 0.1 &   8.5 1.7 & 0.70 &  7.7\nl
FSC 2217+259& 4 & 4.3 & 7.0 0.2 &  43.0 0.5 & 0.28 &  9.7\nl
3C 445  & 4 & 8 & 2.1 0.1 & 142.0 1.0 &$<$0.10&  2.1\nl
\enddata

\tablenotetext{a}{number of bins averaged together for Fig 2 and for the polarized flux in Fig 3}
\tablenotetext{b}{Width of extraction window for spectra}
\tablenotetext{c}{polarization percentage averaged over 5400 - 5600 \AA~(rest frame)}
\tablenotetext{d}{position angle of electric vector, averaged over 5400 - 5600 \AA~(rest frame)}
\tablenotetext{e}{template galaxy fraction at 5500 \AA~}
\tablenotetext{f}{galaxy-corrected polarization at 5500 \AA~(rest)}

\end{deluxetable}


\begin{deluxetable}{lccc}     
\tablewidth{0pt} 
\tablecaption{Broad line visibility}
\tablehead{   & \colhead{High}  & \colhead{Broad} & \colhead{Broad} \nl
          & \colhead{$p_{\rm c}$\tablenotemark{a}} & 
            \colhead{H$\alpha$\tablenotemark{b}} 
          & \colhead{P$\alpha$\tablenotemark{c}}} 
\startdata
QUASAR & & & \nl
\hline
3C 273    &   N\tablenotemark{d}  &   Y   &   Y  \nl
\hline
BLRG & & & \nl
\hline
3C 109    &   Y   &   Y   &\nodata\nl
3C 227    &   Y   &   Y   &\nodata\nl
3C 303    &\nodata&   Y   &   Y   \nl 
3C 382    &   N   &   Y   &\nodata\nl
FSC 2217+259  &   Y   &   Y   &\nodata\nl
3C 445    &   N   &   Y   &\nodata\nl
\hline
NLRG & & & \nl
\hline
3C 33     &   Y   &   Y   &\nodata\nl
3C 105    &   N   &   N   &\nodata\nl
3C 135    &   Y   &   Y   &   N   \nl   
3C 184.1  &\nodata&   N\tablenotemark{e}   &   Y   \nl  
3C 195    &   Y   &   Y   &\nodata\nl
3C 219    &\nodata&   Y\tablenotemark{e}   &   Y   \nl  
3C 223    &\nodata&   Y\tablenotemark{e}   &   Y   \nl  
3C 234    &   Y   &   Y   &   Y   \nl  
3C 236    &  (N)  &\nodata&   N   \nl  
3C 321    &   Y   &   Y   &   N   \nl  
3C 319    &\nodata&   N   &\nodata\nl      
3C 327    &\nodata&\nodata&   N   \nl  
3C 357    &   Y   &   N   &\nodata\nl
Cyg A     &   Y   &   Y   &\nodata\nl
\enddata
\tablenotetext{a}{$p_{\rm c}\ge 3\%$} 
\tablenotetext{b}{Broad H$\alpha$ detected in total or polarized flux}
\tablenotetext{c}{Broad P$\alpha$ detected in total flux (Hill, Goodrich,
                  \& DePoy 1996)}
\tablenotetext{d}{Impey, Malkan, \& Tapia 1989}
\tablenotetext{e}{Hill, Goodrich, \& DePoy 1996}       
\end{deluxetable}


\begin{deluxetable}{lccccc}      
\tablecaption{Radio and Optical Data}
\tablehead{\colhead{Name}& &\colhead{Radio Axis}& &
           \colhead{Optical Pol}& \colhead{$\Delta\theta$} \nl
         &  \colhead{log R \tablenotemark{a}} &
            \colhead{PA $\pm$}& 
            \colhead{ref} & \colhead{$\theta \pm$}&\colhead{(deg)}} 
\startdata
BLRG     &        &           &    &           &   \nl
\hline
3C 109   & -0.58  & 144.0 6   &  1 & 170.5 0.6 & 26\nl
3C 227   & -1.90  &  84.5 7   &  2 &  41.6 0.7 & 43\nl
3C 382   & -1.10  &  52.0\tablenotemark{b}     &  3 & 63.0 0.4 & 11\nl
FSC 2217+259 & -0.40  & 158.0 0.5 &  4 &  43.0 0.5 & 65\nl
3C 445   & -1.41  & 173.0 3.5 &  5 & 142.0 1.0 & 31\nl
\hline
NLRG     &        &           &    &           &   \nl
\hline
3C 33    & -2.25  &  19.5 1   &  6 &  91.0\tablenotemark{c}~ 3.1 & 72\nl
3C 105   & -2.24  & 125.5     &  5 & \nodata & \nodata \nl 
3C 135   & -2.72\tablenotemark{d} & 76.0 3   &  5 & 142.1 4.8 & 66\nl
3C 195   & -1.46  &  19.5 2.5 &  2 & 128.1 1.4 & 71\nl
3C 234   & -1.16  &  67.0 5   &  7 & 156.3 0.4 & 89\nl
3C 321   & -1.78  & 134.5 0.7 &  1 &  61.0\tablenotemark{e}~ 2.0 & 74\nl
3C 357   & -2.21  & 110.0\tablenotemark{f}     &  8 & 24.0 6.6 & 86\nl
Cyg A    & -3.37  & 284.0 2   &  9 &   8.5 1.0 & 84\nl
\enddata
\tablenotetext{a}{Ratio of core to lobe power at $\lambda=6$~cm;
from \cite{zir95} except for 3C 135 (\cite{har98}) and FSC 2217+259 
(\cite{lau97}).}
\tablenotetext{b}{PA from jet in NE lobe}
\tablenotetext{c}{$\theta$ at the nucleus}
\tablenotetext{d}{$\lambda=3.6$~cm}
\tablenotetext{e}{$\theta$ at the SE flux peak}
\tablenotetext{f}{PA from SE lobe}
\tablecomments{References: (1) Baum et al 1988;      
  (2) Morganti et al 1993;
  (3) Black et al 1992;  
  (4) Condon et al 1995;
  (5) Leahy et al 1997;
  (6) Leahy and Pearly 1991; 
  (7) Burns et al 1984;     
  (8) Fanti et al 1986;    
  (9) Carilli et al 1996.}  
    
\end{deluxetable}


\begin{deluxetable}{lccc} 
\tablecaption{Scattering Cone, Outflow, and Radio Jet Sidedness}
\tablehead{ & \colhead{Pol Fan \tablenotemark{a}} & 
              \colhead{Blue Shift \tablenotemark{b}} & 
              \colhead{Radio Jet \tablenotemark{c}}} 
\startdata
3C 33  & NE & NE & \nodata \nl
3C 195 & NE & SW & SW \nl
3C 321 & NW & SE & NW \nl
Cyg A  &  W & W  & W  \nl
\enddata

\tablenotetext{a}{Side with larger polarized fan}
\tablenotetext{b}{Side with blue-shifted narrow emission lines}
\tablenotetext{c}{Side with stronger jet, References in Table 5}

\end{deluxetable}
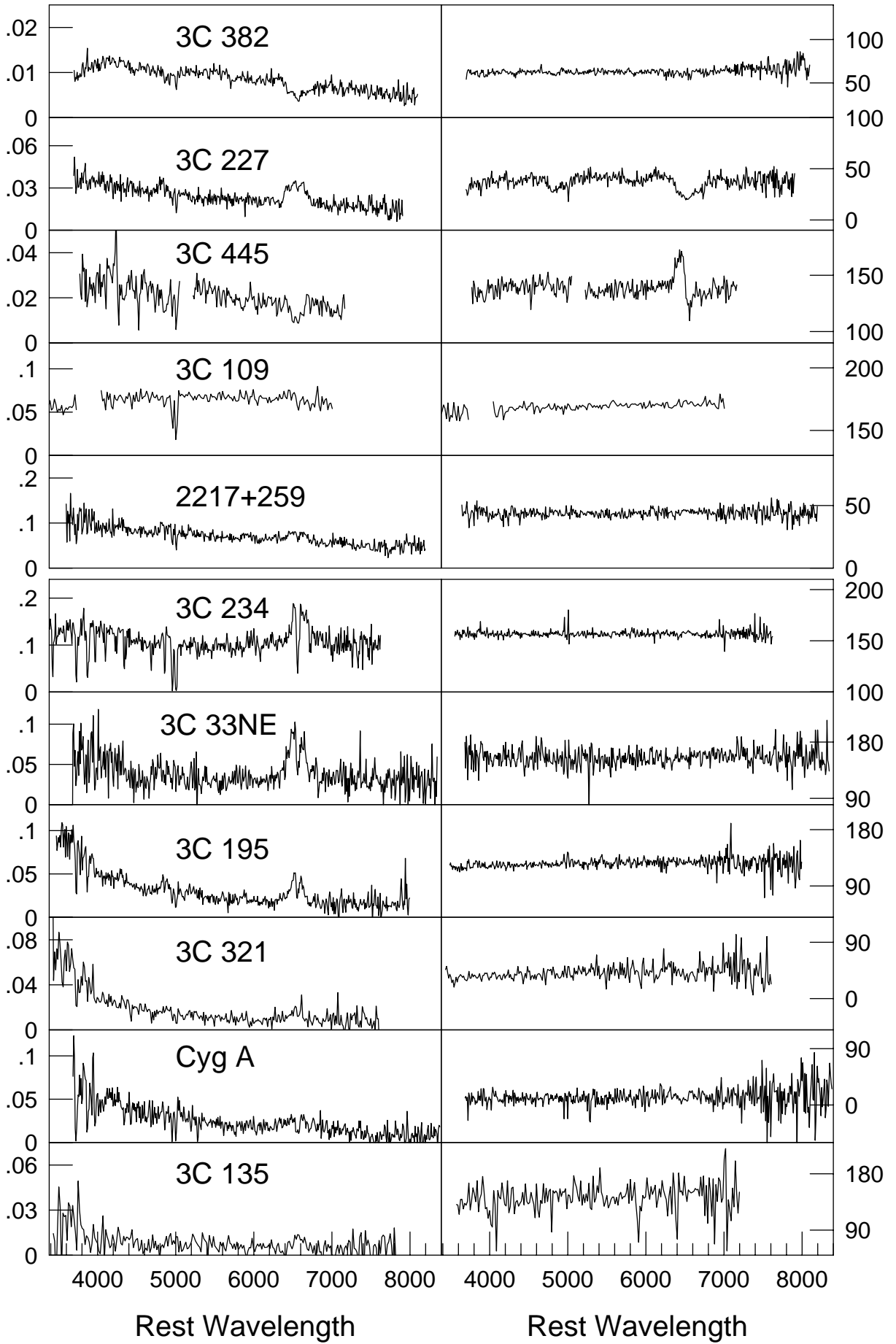
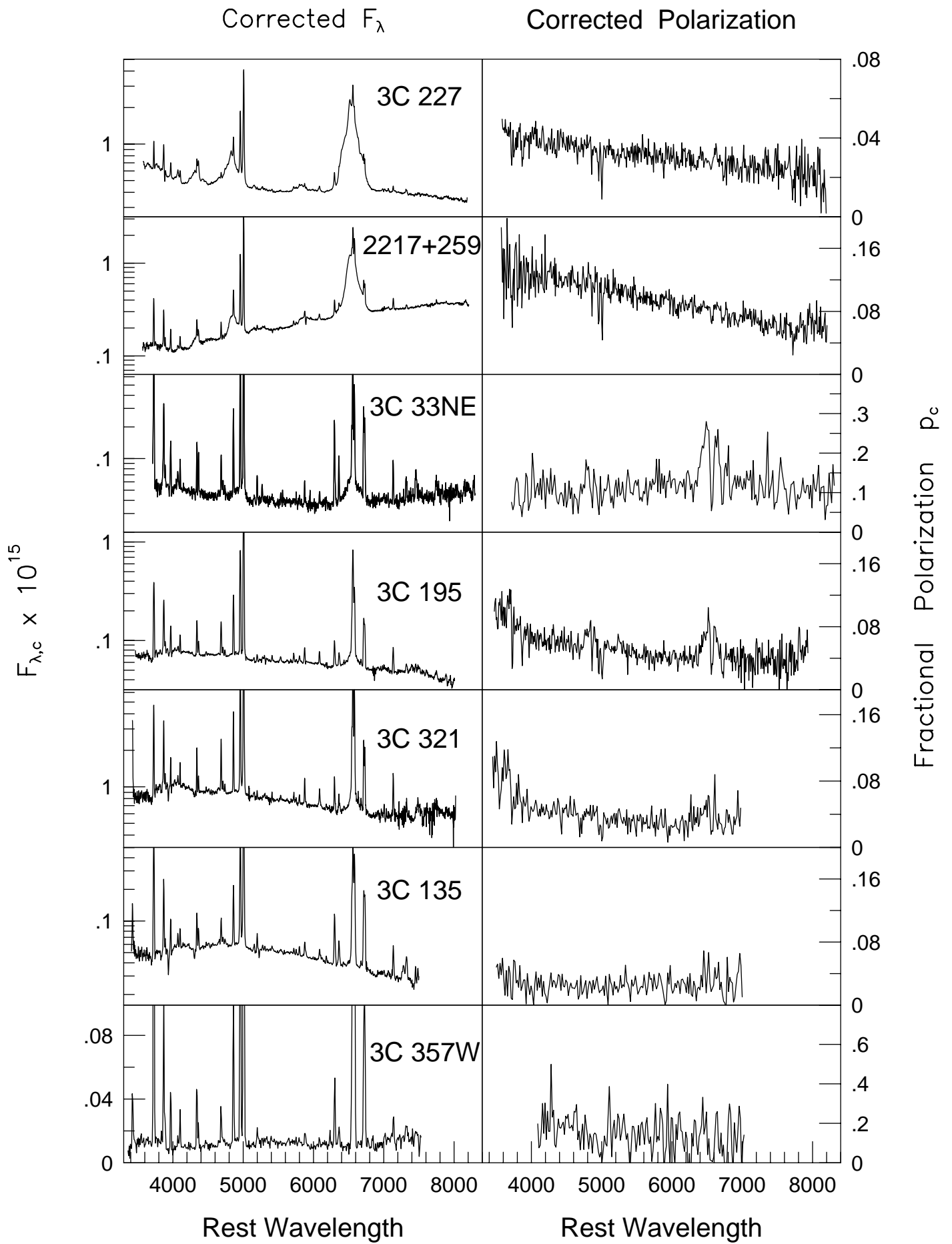
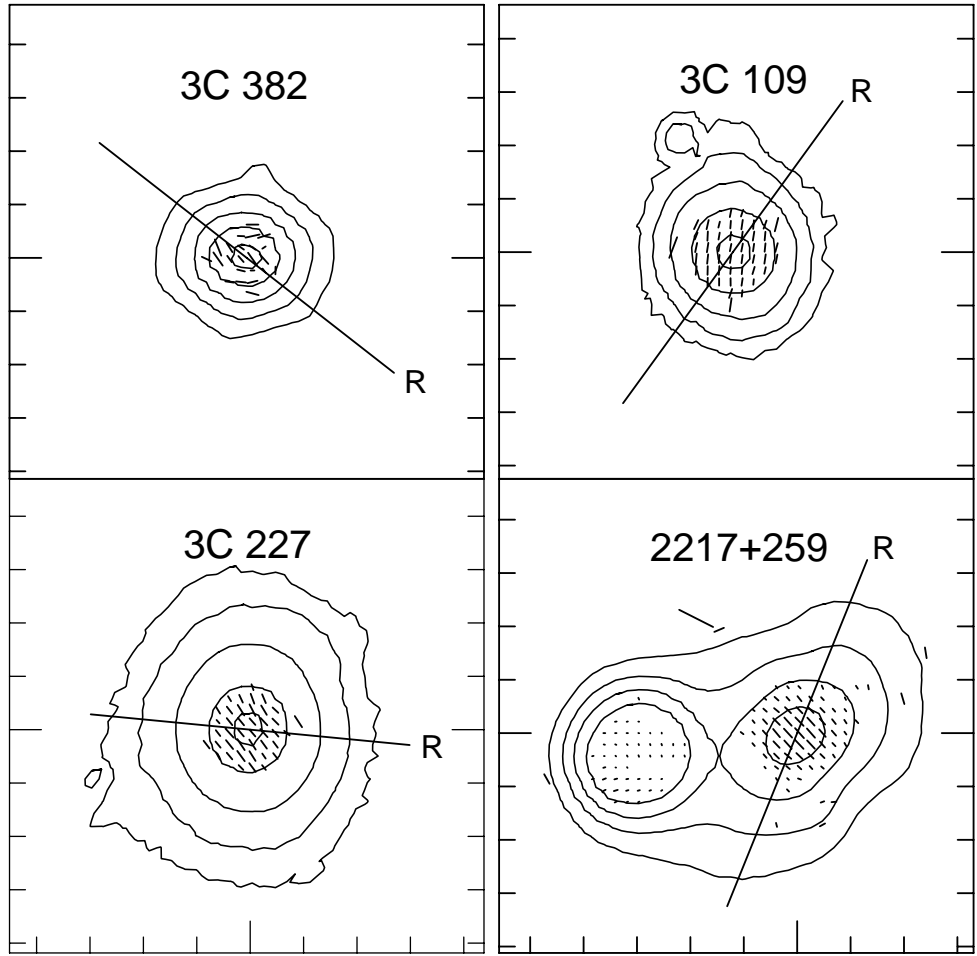
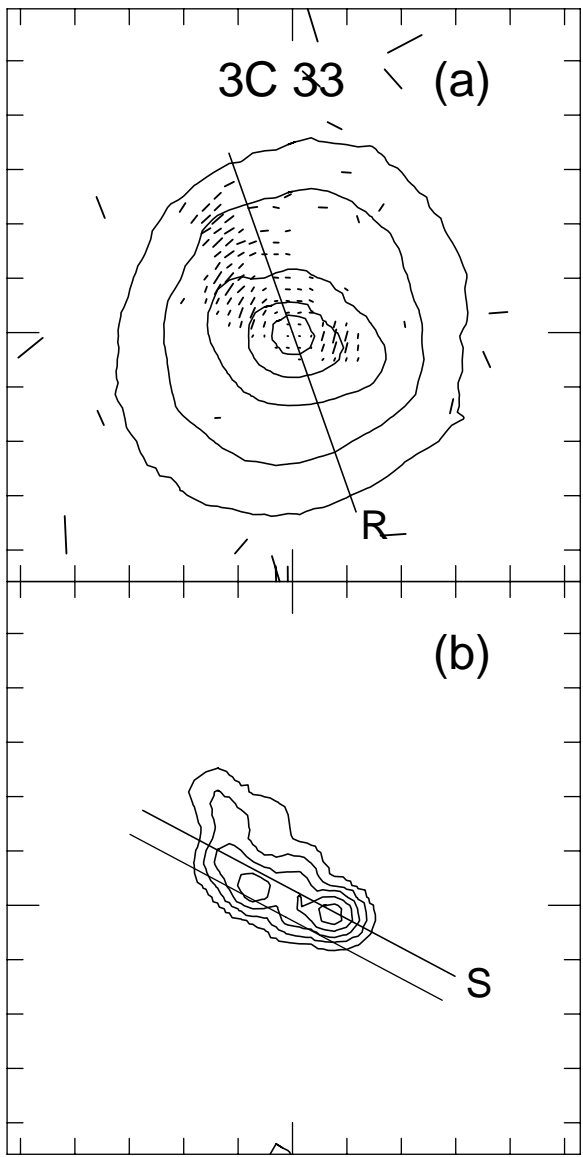
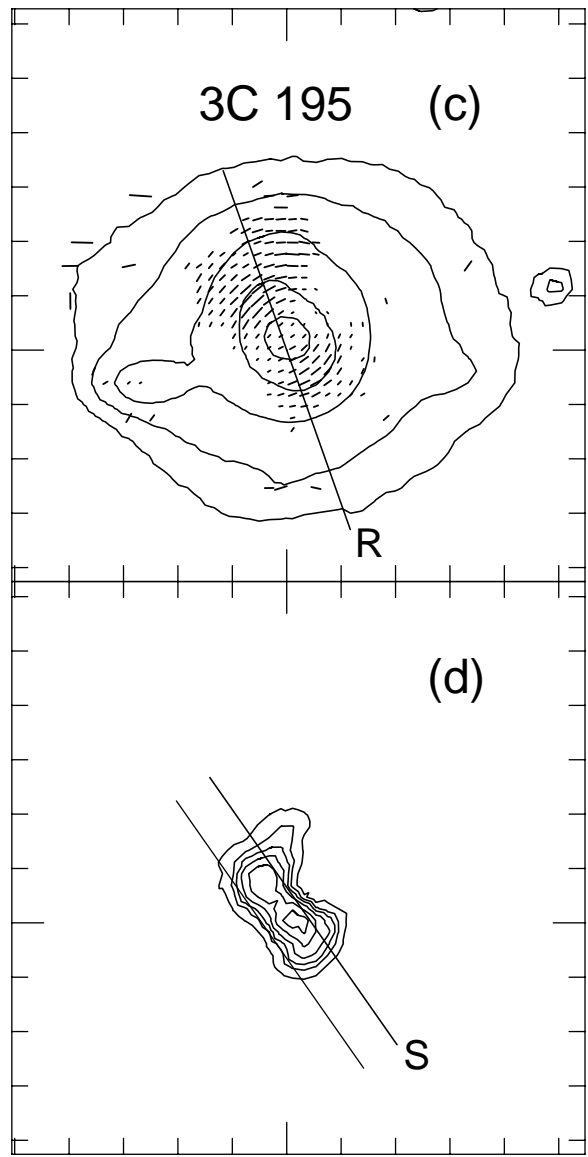
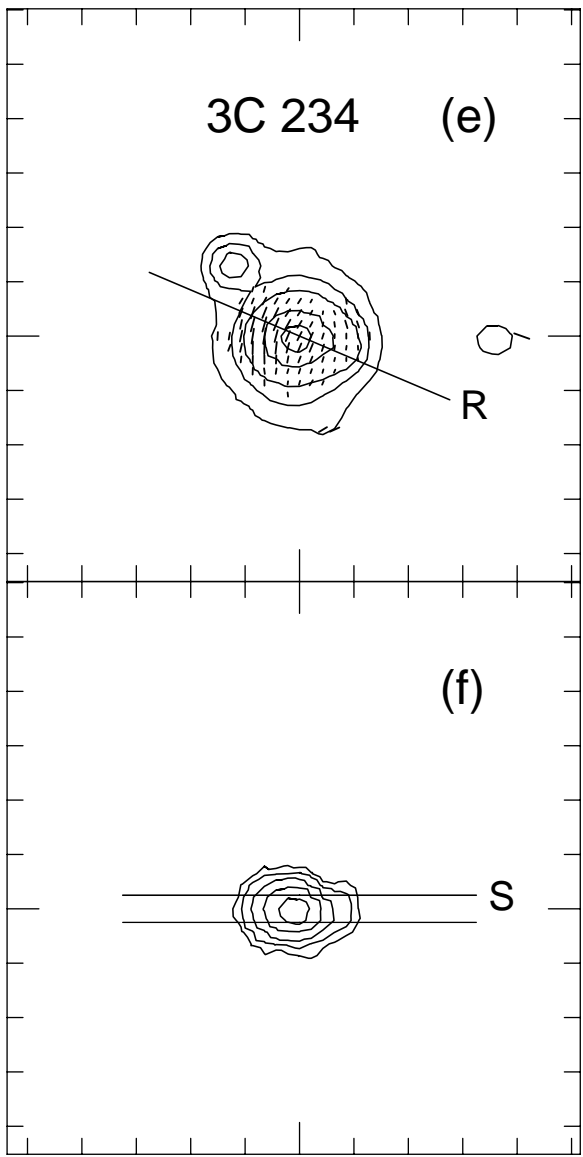
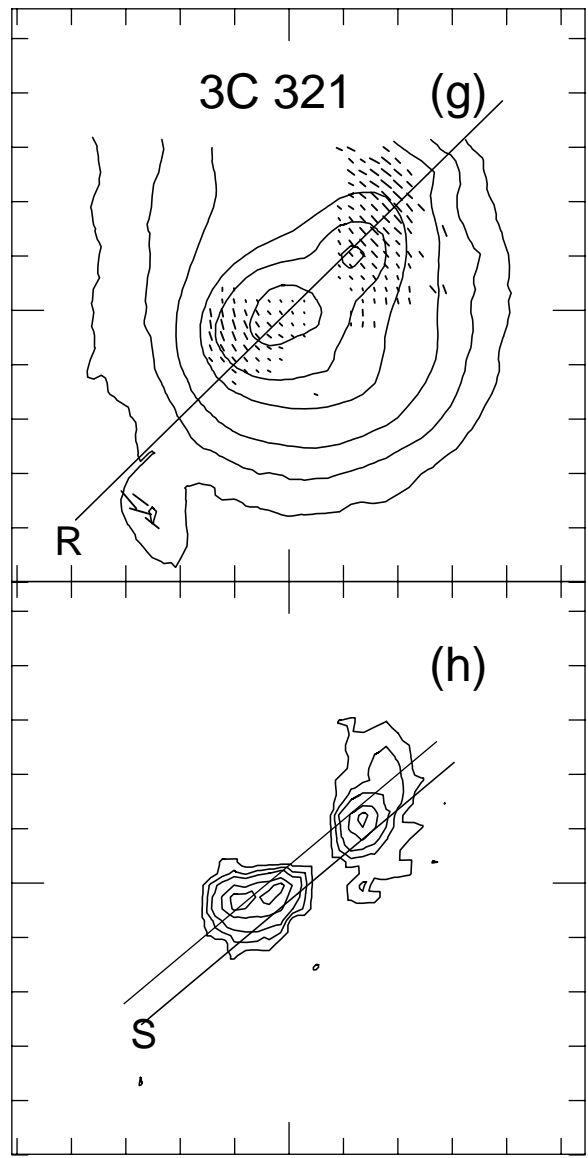
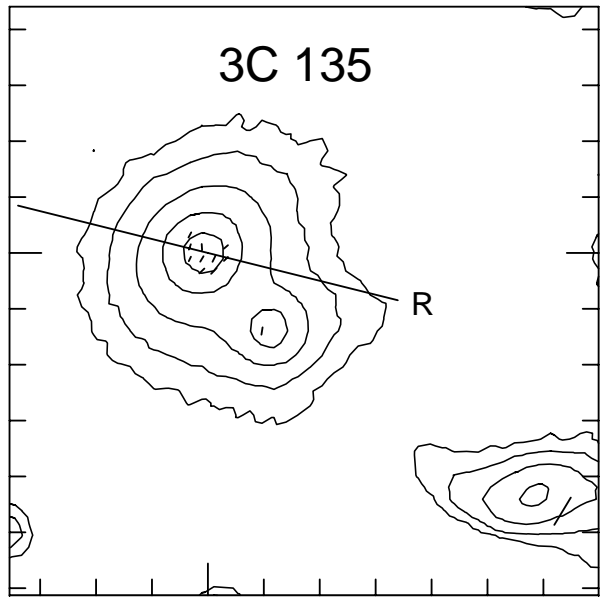
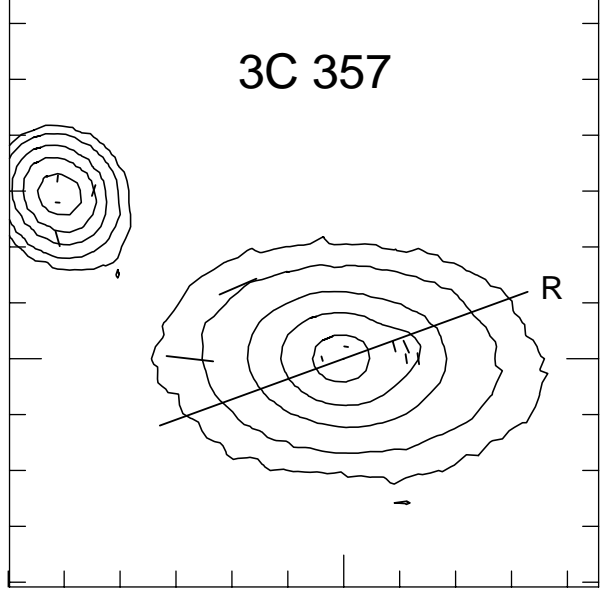
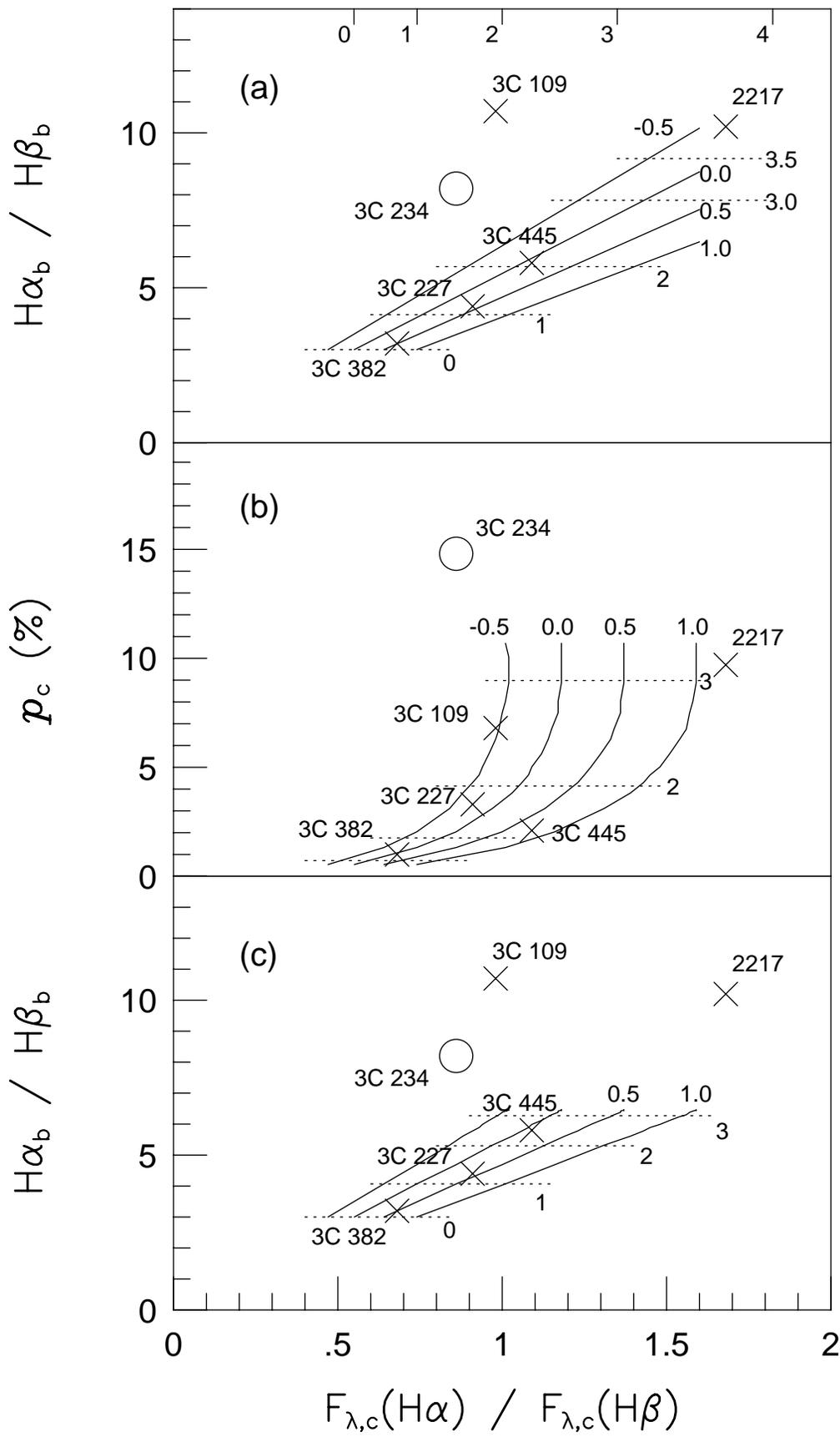
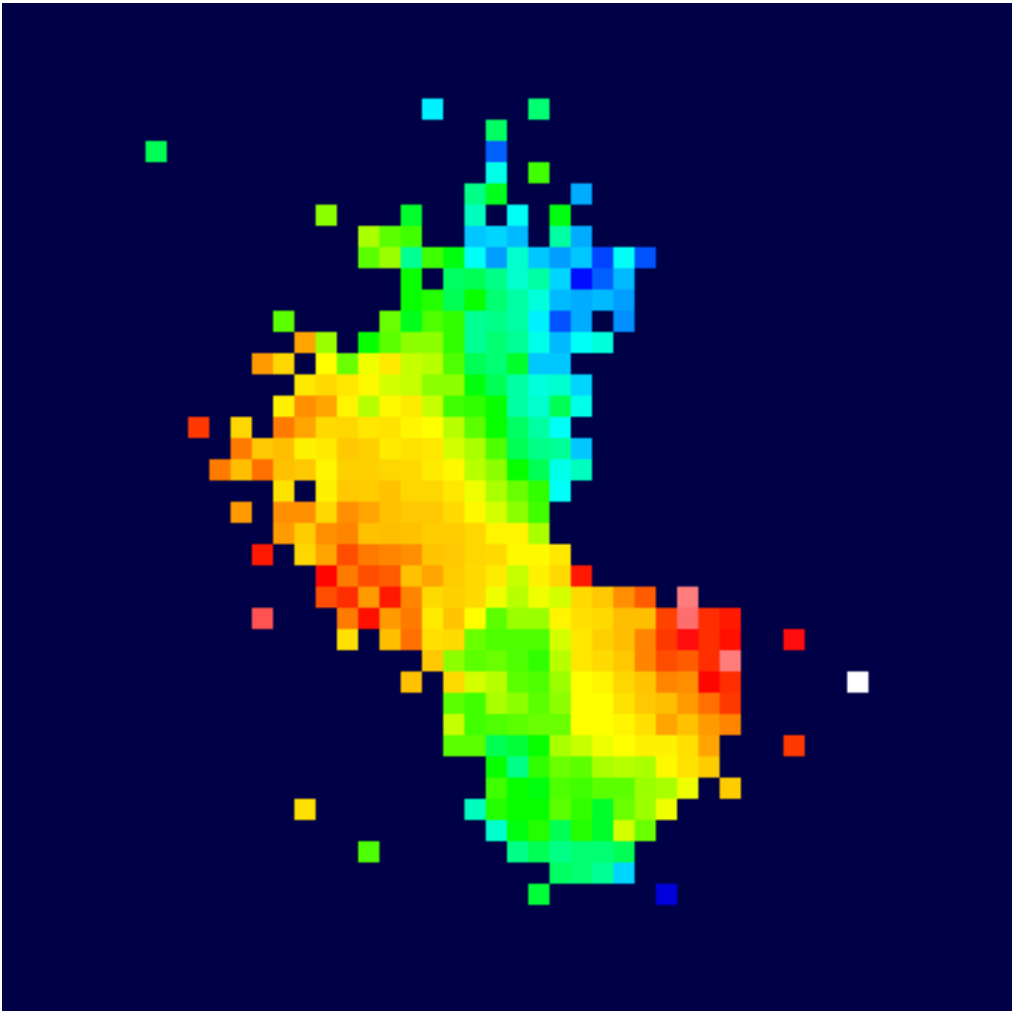
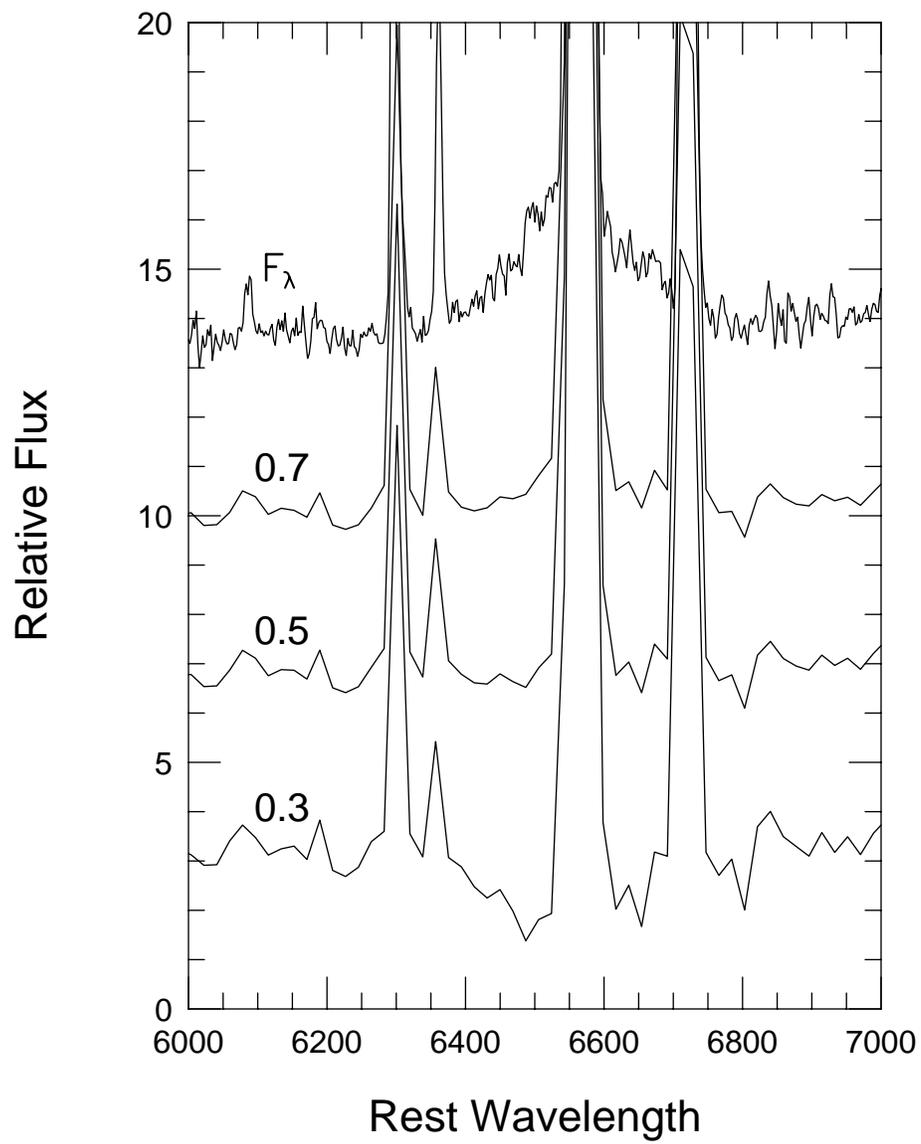
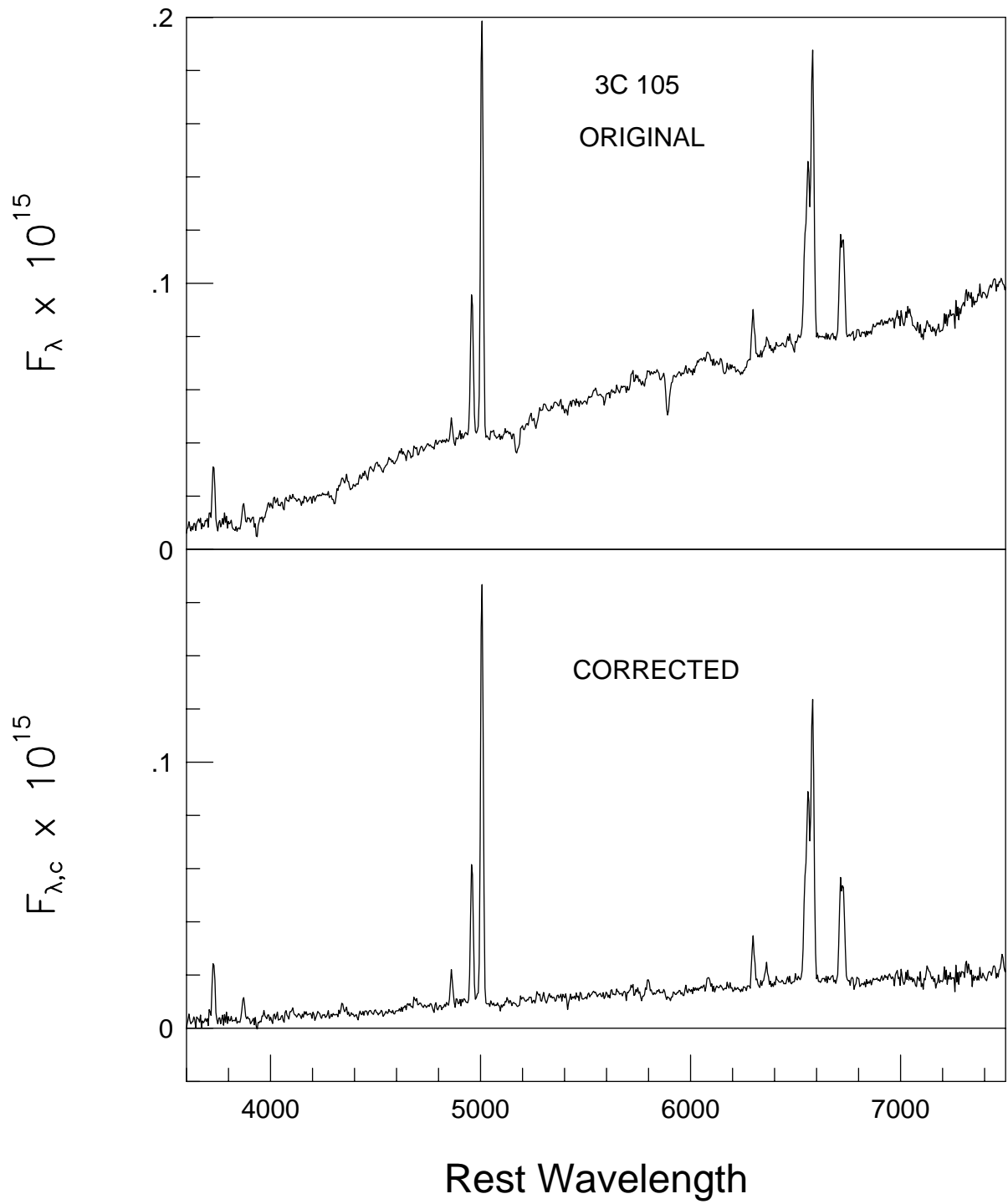
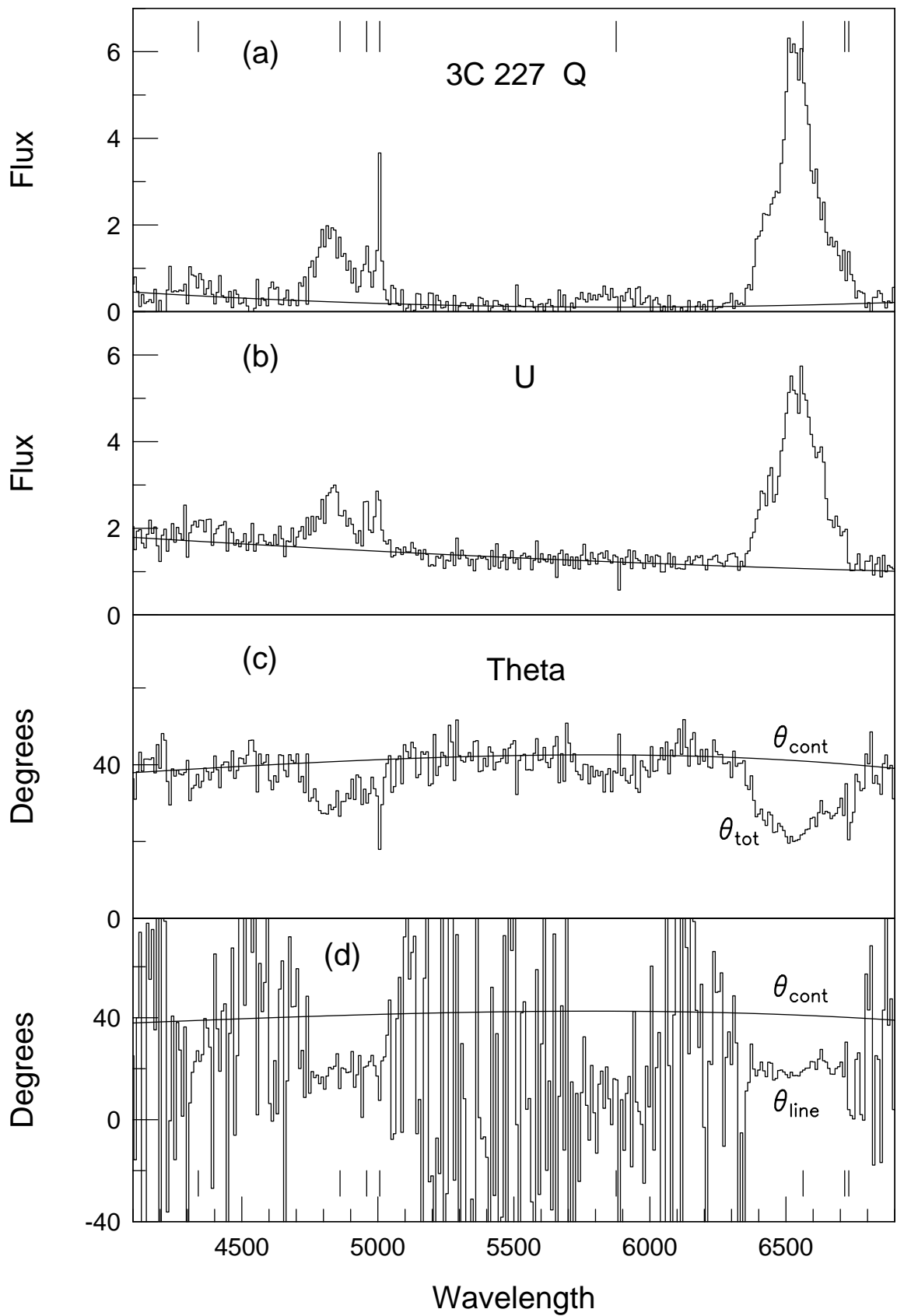
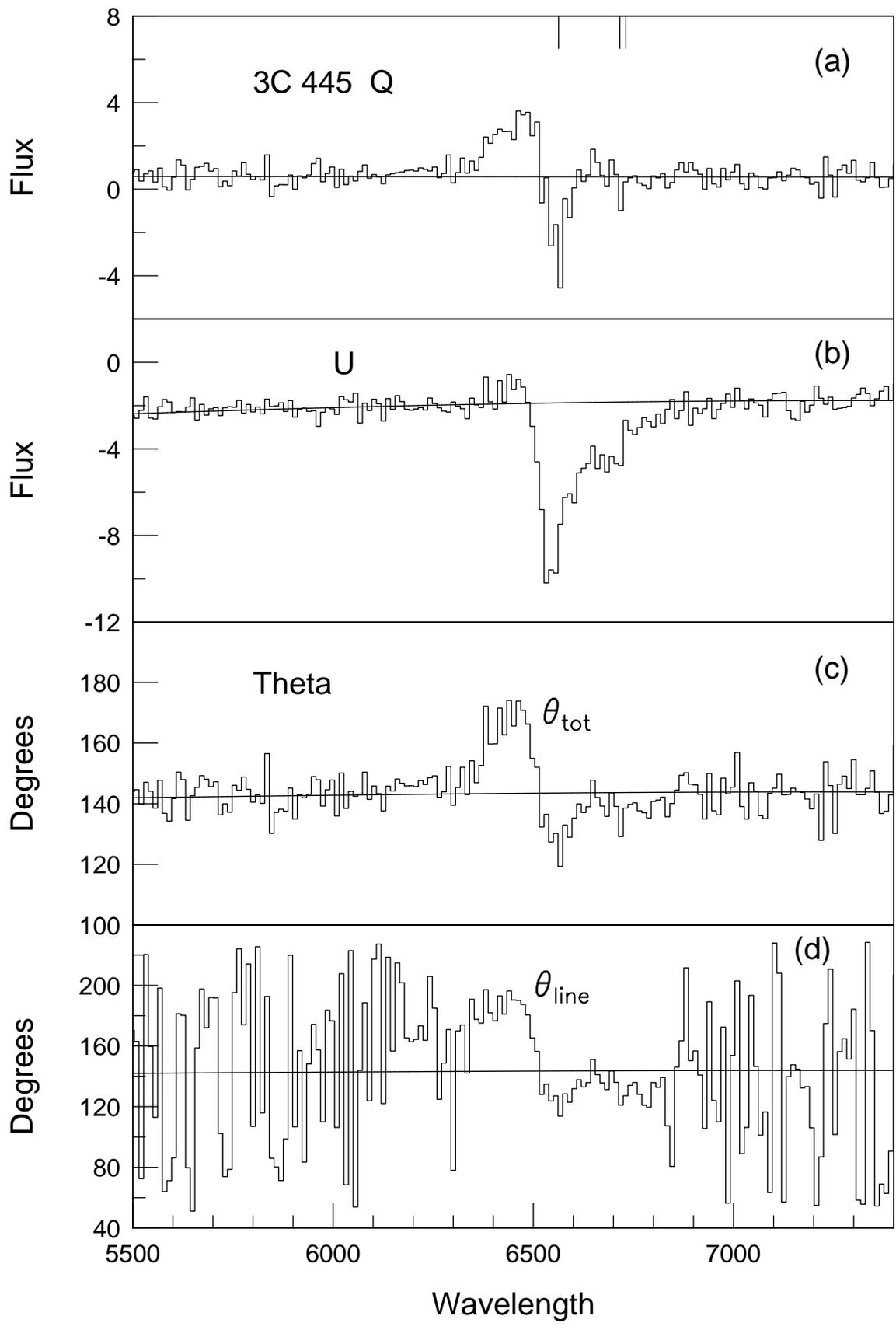
\end{document}